\documentclass{ws-procs975x65}%
\usepackage{amssymb}
\usepackage{amsmath}%
\usepackage{amsfonts}%
\usepackage{graphicx}
\providecommand{\U}[1]{\protect\rule{.1in}{.1in}}
\begin{document}

\title{ON THE COSMOLOGICAL SINGULARITY\footnote{The invited paper for Proceedings of
the XIII Marcel Grossmann Meeting (Stockholm, 2012) by reason of the Marcel
Grossmann Award to V.A. Belinski and I.M. Khalatnikov. }}
\author{\uppercase{Vladimir A. Belinski}}

\address{
ICRANet, Piazzale della Repubblica 10, I--65122 Pescara, Italy;
IHES, 35, Route de Chartres, F-91440 Bures-sur-Yvette, France\\
E-mail: belinski@icra.it\\}%

\begin{abstract}
The long story of the oscillatory approach to the initial cosmological singularity and its more recent incarnation in multidimensional universe models is told.
\end{abstract}%

\bodymatter

\section{The Landau problem and the path to its resolution}

It is known \cite{KK} that in the old days in the 1950s, L.D. Landau expressed
the point of view that the most important problems of theoretical physics were
superconductivity, phase transitions and the cosmological singularity. Today
everyone knows that decisive successes have been achieved in the explanation
of the phenomena of superconductivity (J. Bardeen, L. Cooper and J.
Schrieffer, 1957) and phase transitions (K. Wilson, 1974). We note with
satisfaction that during the last fifty years due to international efforts,
our understanding of the problem of the cosmological singularity has also
reached a deep and sound maturity and it was L.D. Landau himself we have to
thank for the initial spark at the beginning of this impressive development.

The phenomenon of singularities was born in 1922 when A. Friedmann published
his famous cosmological solution for the homogeneous and isotropic Universe.
However, from that time through the late fifties researchers devoted their
attention mainly to the physical processes occurring after the Big Bang and
there were no serious attempts to rigorously analyze the very phenomenon of
the singularity itself. Around the end of 1950s L.D. Landau formulated the
crucial question of whether the cosmological singularity is a generic
phenomenon of general relativity or if it appears only in particular solutions
under special symmetry conditions. He also proposed a way to approach this
problem mathematically. For any system of differential equations one can
define the notion of its general solution which describes the development of
arbitrary initial data, that is, data which is not subject to any special
restrictions. Such a solution should contain a complete set of the arbitrary
functional parameters by which one can specify any possible initial state of
the system. For the gravitational field in vacuum this number is 4, see
Ref.~\refcite{LL}. Of course, due to their mathematical complexity there is no
way to analytically construct the general solution of the gravitational field
equations but for the problem of interest this is not necessary. It would be
enough to find the most general asymptotic expansion of the solution in the
vicinity to the singularity and to count the number of arbitrary functional
parameters it contains. If such asymptotics containing 4 arbitrary
3-dimensional functions can be found, then this will be the proof that a
general solution with a cosmological singularity exists. Without loss of
generality we can choose the synchronous reference system\footnote{The
interval in synchronous system we write in the form: $-ds^{2}=-dt^{2}%
+g_{\alpha\beta}dx^{\alpha}dx^{\beta},$ where $g_{\alpha\beta}$ is positive
definite. We use Greek letters for space-like tensorial indices and Latin
letters in round brackets for the corresponding frame indices. Ordinary
partial derivatives with respect to the coordinates $x^{\alpha}$ we designate
by a comma and the ordinary partial derivative with respect to the synchronous
time $x^{0}=t$ we denote by a dot. We use units in which the Einstein
gravitational constant and the velocity of light are equal to unity.} which is
most suitable for cosmological applications of general relativity and consider
the singular space-like hypersurface (assuming its existence) having equation
$t=0$. Then for each possible type of expansion of the metric tensor with
respect to the infinitesimally small values of time variable, one can perform
the analysis in accordance with Landau's prescription. Of course, this program
entails the knowledge in advance of all possible characters of the expansion
of the metric in the vicinity to the singularity. At the time when these ideas
appeared, only singularities with power law asymptotics for the scale factors
were known and the detailed study of all such cases was performed by E.M.
Lifschitz and I.M. Khalatnikov during the years 1959--1962. They collected and
published their results in 1963 in the comprehensive review of Ref.~\refcite{LK}.

They showed that any solution of the Einstein equations (in vacuum or in a
space filled with a perfect fluid) with a cosmological singularity having
power law asymptotics could not be general. Each solution of this type is
unstable in the sense that if one approaches the singularity, any power law
regime sooner or later will be destroyed under the influence of arbitrary
perturbations. In vacuum the most general solution with a power law
singularity contains only 3 arbitrary functional parameters instead of the 4
which are necessary for the solution to be general. They named this solution
with three arbitrary degrees of freedom the \textit{generalized Kasner
solution} since it represents the inhomogeneous generalization of the
well-known Kasner metric (the vacuum anisotropic homogeneous cosmological
model of Bianchi type I). In the synchronous reference system this solution
describes the anisotropic expansion or contraction of the 3-dimensional space
along three different directions $l_{\alpha}^{\left(  1\right)  },l_{\alpha
}^{(2)},l_{\alpha}^{(3)}$ (Kasner axes) with three different scale factors
$a^{2},b^{2},c^{2}$ corresponding to these directions. The first order terms
of the generalized Kasner solution near the singularity $t=0$ are%

\begin{equation}
g_{\alpha\beta}=a^{2}l_{\alpha}^{(1)}l_{\beta}^{(1)}+b^{2}l_{\alpha}%
^{(2)}l_{\beta}^{(2)}+c^{2}l_{\alpha}^{(3)}l_{\beta}^{(3)},\text{
\ \ }\label{OR1}%
\end{equation}%
\begin{equation}
a^{2}=t^{2p_{\left(  1\right)  }},\text{ }b^{2}=t^{2p_{\left(  2\right)  }%
},\text{ }c^{2}=t^{2p_{\left(  3\right)  }},\label{OR1-1}%
\end{equation}
where the exponents $p_{\left(  a\right)  }$ and vectors $l_{\alpha}^{(a)}$
depend only on three-dimensional space coordinates $x^{\alpha}$. The
generalized Kasner exponents $p_{\left(  a\right)  }$ have to satisfy two relations:%

\begin{equation}
p_{\left(  1\right)  }+p_{\left(  2\right)  }+p_{\left(  3\right)  }=1,\text{
\ \ }p_{\left(  1\right)  }^{2}+p_{\left(  2\right)  }^{2}+p_{\left(
3\right)  }^{2}=1.\label{OR2}%
\end{equation}
This approximation comes from gravitational field equations $R_{00}=0$ and
$R_{\alpha\beta}=0$ when one only keeps terms with time derivatives in them,
neglecting all terms containing spatial derivatives (the equations
$R_{0\alpha}=0$ are of no interest since they only put three constraints on
the arbitrary 3-dimensional functions $p_{\left(  a\right)  },l_{\alpha}^{(a)}
$ and play no role in the dynamics). From the relations (\ref{OR2}) it follows
that one of the exponents is negative and other two are positive.
Consequently, when one approaches the singularity $t\rightarrow0$ the
3-dimensional space expands in one direction and contracts in two others.

Without loss of generality we can choose $p_{\left(  1\right)  }$ to be
negative and in this case the generalized Kasner asymptotics will be valid up
to the singularity if and only if the vector $l_{\alpha}^{(1)}$ (namely that
one corresponding to the direction of unlimited expansion near the
singularity) is subject to the following additional constraint (additional
with respect to the 3 natural constraints following from the equations
$R_{0\alpha}=0$):%
\begin{equation}
l_{(2)}^{\mu}l_{(3)}^{\nu}[l_{\mu,\nu}^{(1)}-l_{\nu,\mu}^{(1)}]=0,\label{OR3}%
\end{equation}
where the vectors $l_{(a)}^{\alpha}$ are inverse to the $l_{\alpha}^{(a)},$
that is $l_{(a)}^{\alpha}l_{\beta}^{(a)}=\delta_{\beta}^{\alpha}$ and
$l_{(b)}^{\alpha}l_{\alpha}^{(a)}=\delta_{\left(  b\right)  }^{\left(
a\right)  }.$ This condition comes from the fact that without it the Kasner
dynamics with $p_{\left(  1\right)  }<0$ ceases to be valid in the limit
$t\rightarrow0$ because the terms of the next approximation to the metric
tensor (\ref{OR1})--(\ref{OR2}) instead of being negligible diverge relative
to the terms of first order. This divergence is due the influence of those
terms in the equations $R_{\alpha\beta}=0$ which contain the quantity
$l_{(2)}^{\mu}l_{(3)}^{\nu}[l_{\mu,\nu}^{(1)}-l_{\nu,\mu}^{(1)}]$ and its
space derivatives. Then if this quantity does not vanish, the original
assumption of the dominating role of the time derivative terms is violated.
However, it turns out that all these divergences can be removed by only one
additional constraint \textbf{(\ref{OR3})}. After that all higher order
corrections to the generalized Kasner asymptotics near the singularity will be
infinitesimally small, which justifies the validity of the approximation
(\ref{OR1})--(\ref{OR2}). All these higher order corrections can be calculated
from the Einstein equations in a unique way and they do not contain any new
arbitrary 3-dimensional functions apart from those already presented in the
solution of the first approximation.

The qualitative character of the generalized Kasner solution remains the same
also in the presence of a perfect fluid with any equation of state except the
stiff matter $p=\varepsilon$ case (for this special Zeldovich type of equation
of state see below). The presence of a perfect fluid does not change the
character of the leading asymptotics (\ref{OR1})--(\ref{OR2}) of the solution
and matter can be ``inserted" in the solution together with its 4 additional
arbitrary three dimensional functions (one for the energy density and three
for the velocity).

Consequently their work \cite{LK} showed that the most general cosmological
solution (for vacuum and standard perfect fluid cases) with power law
asymptotics has one degree of freedom less with respect to the general
solution and the additional condition (\ref{OR3}) is responsible for this.

Of course, from this result one cannot draw any conclusion about the existence
or non-existence of a general solution with a singularity. It only means that
the general solution cannot have a singularity of power law character, but
this does not exclude the possibility that a general solution with a
singularity of some different yet unknown type does exist. The natural way to
try to resolve the problem was to look at what will happen to the generalized
Kasner solution after (towards the singularity $t\rightarrow0$) the terms in
the gravitational equations containing the non-zero quantity $l_{(2)}^{\mu
}l_{(3)}^{\nu}[l_{\mu,\nu}^{(1)}-l_{\nu,\mu}^{(1)}]$ begin to destroy the
Kasner regime. In physics the missions of this kind usually are highly
non-trivial but in general relativity, in spite of the complexity of the
Einstein equations, the final fate of the gravitational field near a
singularity can be described almost completely. This was done mainly during
1963--1968 by V. Belinski, I. Khalatnikov and E. Lifshitz (BKL) who showed
that after removing the artificial additional condition (\ref{OR3}) the
solution develops to the \textit{generic} solution of the gravitational field
equations possessing a real curvature singularity of an essentially new type
of chaotic oscillatory character instead of a smooth power law behavior. The
detailed exposition of the BKL results can be found in their basic articles
\cite{BK1,BK2,LLK,BKL1,BKR} and in two review papers \cite{BKL2,BKL3}.

In order to avoid misunderstandings let's stress that by a cosmological
singularity we mean a singularity in time, where the singular manifold is
space-like, and the curvature invariants together with the invariant
characteristics of any matter fields (like energy density) diverge on this
manifold. In 1965 Roger Penrose \cite{Pen} proved an important theorem, saying
that under some conditions the appearance of incomplete geodesics in
space-time is unavoidable. This is also a singularity but of a different type
since, in general, incompleteness does not means that invariants diverge. In
addition the theorem can say nothing about the analytical structure of the
fields near the points where geodesics terminate. Thus Penrose's result was
not of any direct help for us, nevertheless it stimulated our search. In any
case it is true to say that the BKL approach and the Penrose theorem
elucidates two sides of the phenomenon but the links between them are still
far from understood. This is because the BKL approach studies the asymptotics
in the vicinity of the the singularity and Penrose's theorem has to do with
global properties of the space-time.

Following idea of the BKL analysis one should include in the equations of the
main approximation not only the time derivatives but also those spatial
gradients which are responsible for the instability of the Kasner dynamics
near the singularity. If one approaches the singularity starting with the
metric (\ref{OR1})--(\ref{OR2}) with $p_{\left(  1\right)  }<0$ then it is
necessary to take into account first of all those terms in the equations which
are present due to the non-zero quantity $l_{(2)}^{\mu}l_{(3)}^{\nu}%
[l_{\mu,\nu}^{(1)}-l_{\nu,\mu}^{(1)}]$ . It is remarkable that the influence
of these terms can be clarified exactly and result shows that their action
transforms the initial Kasner regime again to another Kasner regime but with
new 3-dimensional functional parameters $\acute{p}_{\left(  a\right)  }%
,\acute{l}_{\alpha}^{(a)}.$ These new parameters can be expressed uniquely in
terms of their initial values $p_{\left(  a\right)  },l_{\alpha}^{(a)}.$ The
crucial point is that the new exponents $\acute{p}_{\left(  a\right)  }$ still
satisfy the same Kasner relations (\ref{OR2}) and one of them again will be
negative and two others positive. If initially we started with the asymptotics
(\ref{OR1})--(\ref{OR2}) with $p_{\left(  1\right)  }<0$ then at the new
\textit{Kasner} \textit{epoch} the negative exponent jumps from the scale
factor $a^{2}$ to the scale factor $b^{2}$ or $c^{2}$ (to that one which had
the smallest of two positive Kasner exponents at the initial epoch). Namely
$a^{2}$ in fact does not diverge as $t\rightarrow0$ but at some critical time
reaches a maximum and then starts to decrease while one of the factors $b^{2}$
or $c^{2}$ starts to increase instead. Exactly what happens is the following.
If we start the initial Kasner epoch with $p_{\left(  1\right)  }<p_{\left(
2\right)  }<p_{\left(  3\right)  }$ (i.e. $p_{\left(  1\right)  }<0$ and
$p_{\left(  3\right)  }>p_{\left(  2\right)  }>0$) by the asymptotics%
\begin{equation}
a^{2}=a_{0}^{2}\,t^{2p_{\left(  1\right)  }},\text{ }b^{2}=b_{0}%
^{2}\,t^{2p_{\left(  2\right)  }},\text{ }c^{2}=c_{0}^{2}\,t^{2p_{\left(
3\right)  }},\label{OR4}%
\end{equation}
then after the aforementioned critical time the asymptotics of the scale
factors changes to%
\begin{equation}
a^{2}=\tilde{a}_{0}^{2}\,t^{-\frac{2p_{\left(  1\right)  }}{1+2p_{\left(
1\right)  }}},\text{ }b^{2}=\tilde{b}_{0}^{2}\,t^{\frac{2\left[  p_{\left(
2\right)  }+2p_{\left(  1\right)  }\right]  }{1+2p_{\left(  1\right)  }}%
},\text{ }c^{2}=\tilde{c}_{0}^{2}\,t^{\frac{2\left[  p_{\left(  3\right)
}+2p_{\left(  1\right)  }\right]  }{1+2p_{\left(  1\right)  }}},\label{OR5}%
\end{equation}
where $a_{0}^{2},$ $b_{0}^{2},$ $c_{0}^{2}$ are the arbitrary 3-dimensional
functions and three new functional parameters $\tilde{a}_{0}^{2},$ $\tilde
{b}_{0}^{2},$ $\tilde{c}_{0}^{2}$ are expressed in terms of $a_{0}^{2},$
$b_{0}^{2},$ $c_{0}^{2},$ $p_{\left(  a\right)  }$ and quantity $l_{(2)}^{\mu
}l_{(3)}^{\nu}[l_{\mu,\nu}^{(1)}-l_{\nu,\mu}^{(1)}]$. Now $b^{2}$ increases
because $p_{\left(  2\right)  }+2p_{\left(  1\right)  }<0.$

This means that in the general case when initially all three quantities
\begin{equation}
\lambda=l_{(2)}^{\mu}l_{(3)}^{\nu}[l_{\mu,\nu}^{(1)}-l_{\nu,\mu}^{(1)}],\text{
}\mu=l_{(3)}^{\mu}l_{(1)}^{\nu}[l_{\mu,\nu}^{(2)}-l_{\nu,\mu}^{(2)}],\text{
}\nu=l_{(1)}^{\mu}l_{(2)}^{\nu}[l_{\mu,\nu}^{(3)}-l_{\nu,\mu}^{(3)}%
]\label{OR3-1}%
\end{equation}
are non-zero, at the second Kasner epoch the system will find itself
qualitatively in the same state as before and the new transition of the same
kind to the third Kasner epoch will take place and so on. This process of the
changing of Kasner epochs produces oscillations of the scale factors. The BKL
analysis shows that these oscillations never terminate up to the singular
point $t=0$ which is the point of their condensation. In synchronous time the
frequencies of these oscillations tend to infinity and the number of epochs
between any instant $t>0$ and singularity $t=0$ is infinite. The duration of
these epochs tends to zero and transitions between them take a very short time
compared to their durations. When one approaches the singularity the values of
all the scale factors, although oscillating, tend to zero because their
successive maxima tend to zero. The determinant of the metric tensor
$g_{\alpha\beta}$ tends to zero monotonically and all invariants of the
Riemann tensor diverge at the point $t=0.$ The solution of this type is
structurally stable and generic since it is free of any additional constraints
on its 3-dimensional parameters apart of those natural ones which are dictated
by the conventional initial conditions for the gravitational field equations.
This singularity is essentially of a new type with respect to all those
previously known because no power law behavior can be ascribed to it.
Moreover, in the asymptotic vicinity of the singularity $t\rightarrow0$ this
oscillatory regime becomes stochastic.

\section{On the stochasticity of the gravitational field near a cosmological
singularity}

To see the stochasticity of the oscillatory cosmological singularity let's
mention first that BKL showed that in the course of approaching the singular
point $t=0$, alternating Kasner epochs are grouped into successive series
called \textit{eras.} During each era two of the scale factors oscillate and
the third one monotonically decreases. At each new era one of the factors
which in a preceding era was oscillating starts to decrease monotonically and
other two start to oscillate against each other and so on. From epoch to epoch
inside each era the Kasner exponents change their places relative to the
Kasner directions and change their numerical values. It is important that at
each new epoch the new exponents $\acute{p}_{\left(  a\right)  }$ are simple
elementary functions on their previous values $p_{\left(  a\right)  }$ only,
that is $\acute{p}_{\left(  a\right)  }$ do not depend on the old and new
directional vectors of the Kasner axis. This means that the whole evolution of
the numerical values of Kasner exponents can be described by the
one-dimensional map since due to the relations (\ref{OR2}) all three exponents
can be expressed in terms of only one free parameter. Let's look on the
evolution of the numerical values of the exponents ignoring the question to
which direction of the 3-dimensional space we should relate that or another
exponent at each Kasner epoch. To do this it is convenient to introduce the
ordered set $p_{\left(  1\right)  }<p_{\left(  2\right)  }<p_{\left(
3\right)  }$ of three Kasner exponents by the following parametric
representation (which automatically meets the requirements (\ref{OR2})):
\begin{equation}
p_{\left(  1\right)  }=\frac{-u}{1+u+u^{2}},\text{ }p_{\left(  2\right)
}=\frac{1+u}{1+u+u^{2}},\text{ }p_{\left(  3\right)  }=\frac{u(1+u)}%
{1+u+u^{2}},\label{OR3-2}%
\end{equation}
where the functional parameter $u(x^{1},x^{2},x^{3})$ takes values in the
region $u\geqslant1.$ The exponents appearing in the metric (\ref{OR1}) for
each Kasner epoch follow from this ordered set by permutations. It turns out
that if we start the evolution towards the singularity with some initial value
of parameter $u^{(1)}>1$ then the evolution of the numerical values of the
exponents (\ref{OR3-2}) at the first era corresponds to the sequence
$u^{(1)},u^{(1)}-1,u^{(1)}-2,...$(this can be derived from the law
(\ref{OR4})--(\ref{OR5}) governing the change of Kasner exponents). Because
the range of parameter $u$ is fixed to be greater than unity we can follow
such a sequence up to the value $u^{(1)}-[u^{(1)}]+1$, where $[u^{(1)}]$ means
the integer value of $u^{(1)}.$ The last Kasner epoch of the first era
corresponds to the parameter $u^{(1)}-[u^{(1)}]+1=1+x^{(1)}$ where $x^{(1)}<1
$ and after this a new era should start with a new initial epoch corresponding
to the parameter $u$ which would be equal to $x^{(1)}$ following the rule
$u\rightarrow u-1$. However, the representation (\ref{OR3-2}) is invariant
under the transformation $u\rightarrow1/u$ (plus permutations of the positive
exponents $p_{\left(  2\right)  }$ and $p_{\left(  3\right)  }$ but, as we
stressed already, we are doing analysis ignoring any permutations of the
exponents). Due to this freedom we can arrange the initial parameter $u^{(2)}$
on the first epoch of the new era again to be greater then unity performing
transformation $u^{(2)}=1/x^{(1)}.$ Now the evolution of Kasner exponents
during the second era can be described by the same sequence $u^{(2)}%
,u^{(2)}-1,u^{(2)}-2,...$ up until the final epoch in this second era when the
parameter $u$ will be reduced to the value $u^{(2)}-\left[  u^{(2)}\right]  $
$+1=1+x^{(2)}$ with $x^{(2)}<1$ and so on. It is easy to see that in the heart
of this infinite process lie the one-dimensional discrete map
\begin{equation}
x^{\left(  s+1\right)  }=\left\{  \frac{1}{x^{\left(  s\right)  }}\right\}
,\label{OR3-3}%
\end{equation}
where curly brackets mean the fractional part of a number and by the index $s
$ we enumerate the eras. The number $x^{\left(  s\right)  }$ from the interval
$\left(  0,1\right)  $ represents the fractional part of the value $u^{\left(
s\right)  }$ of the parameter $u$ at the first Kasner epoch of the era with
numerical index $s$.

Mathematicians had studied this map (\ref{OR3-3}) long ago and it was known to
be stochastic in the sense that in the limit $s\rightarrow\infty$, any initial
probability distribution $w^{\left(  1\right)  }\left(  x^{\left(  1\right)
}\right)  $ for the variable $x^{\left(  1\right)  }$ (no matter how sharp)
will develop into the stationary (invariant) distribution $w\left(  x\right)
$ independent of the initial data. This is characteristic of strong
stochasticity. This limiting stationary distribution (normalized in the
interval $\left(  0,1\right)  $) is:%
\begin{equation}
w\left(  x\right)  =\frac{1}{\left(  1+x\right)  \ln2}.\label{OR3-4}%
\end{equation}

The map (\ref{OR3-3}) had been introduced and analyzed by Carl F. Gauss around
200 years ago in his study of number theory and continued fractions. He also
obtained the distribution (\ref{OR3-4}) but never published its derivation
\cite{Kh}. A rigorous derivation (relatively complicated) was constructed only
in 1928 by the Russian mathematician R.O. Kuzmin \cite{Kuz} and in
contemporary literature this result is often cited as the Gauss-Kuzmin theorem.

The fact that the Gauss map (\ref{OR3-3}) is the source of stochasticity in
the cosmological oscillatory regime near a singularity was first pointed out
in Ref.~\refcite{LLK}. In this paper the statistical distributions and mean
values for other characteristics of the regime were also found and a way was
proposed to interpret them (the question of physical interpretation is
important because distributions which follow from the Gauss-Kuzmin formula
(\ref{OR3-4}) are unstable, namely fluctuations around mean values are not small).

After the discussion of the stochasticity of the oscillatory regime in
\cite{LLK}, a vast literature appeared dedicated to the further development of
the theory of chaos near the cosmological singularity. The principal steps of
this development (in chronological order) are the following.

In 1972 D. Chitre \cite{Chitre} established that the spatially homogenous
Bianchi type IX system of ordinary differential equations following from the
vacuum Einstein equations is equivalent to a billiard on the Lobachevsky
plane. It is well known that this kind of system has stochastic properties
(see the next section ``Cosmological billiard").

In 1973 appeared important results of O. Bogoyavlenskii and S. Novikov
\cite{BN} who, using the qualitative theory of differential equations (the
theory of dynamical systems), discovered the existence in the Bianchi IX phase
space of an attractor consisting of a dense set of periodic trajectories.
Today we know that this is one of the characteristic features of the strange
attractor signifying the presence of chaos. This approach has an impressive
contemporary continuation (technically quite different) in the work of Refs.~\refcite{UH1, UH2}.

In 1982 J. Barrow \cite{Bar} showed that one can attribute positive
Kolmogorov-Sinai entropy to the stochastic evolution of the parameter $u$ of
Eq.~(\ref{OR3-2}). He calculated this entropy and proved that it is not zero.
Since the Kolmogorov-Sinai entropy can be defined as the sum of the positive
Lyapunov indices, his result means that at least one of these indices is
positive, but this is enough for the appearance of stochasticity.

In 1983 D. Chernoff and J. Barrow \cite{CB} studied the stochastization of the
Bianchi type IX model in general (not only for parameter $u$). They considered
the complete set of Bianchi IX phase space variables (scale factors $a,b,c$
and its first derivatives $\dot{a},\dot{b},\dot{c}$) and separated the
non-stochastic monotonous degrees of freedom (like the determinant of the
metric tensor) from those two (namely the parameter $u$ and ratio $\ln b/\ln
c$) which are stochastic. For these two they obtained a two-dimensional
stochastic map which can be reduced to the already known Baker map, containing
Kolmogorov-Sinai and topological entropy, ergodicity and mixing. For this
two-dimensional map they also found the asymptotic stationary distribution.

In 1985 I.M. Khalatnikov, E.M. Lifshitz, Ya.G. Sinai, K.M. Khanin and L.N.
Shchur \cite{KLKSS} completed the construction of the statistical theory of
the Bianchi type IX oscillatory regime, the basics of which had been worked
out in Ref.~\refcite{LLK}, but those early results (first of all the stable
distributions for the scale factors $a^{2},b^{2},c^{2}$ and their stable mean
values) had been obtained with the help of some approximations, the validity
of which was not evident at first glance. In Ref.~\refcite{KLKSS} the
confirmation of the validity of these approximations and construction of the
corresponding exact approach was given.

Then a long period (1988--1995) of debate followed during which various
authors (for a review see the book Ref.~\refcite{Hob}) expressed doubts about
the statement that Bianchi type IX dynamics is really chaotic. At that time
some authors even expressed their opinion that the Bianchi type IX model might
be an integrable system. Such a conjecture was erroneous, of course. The
discussion had its origin in the fact that the Lyapunov indices are not
invariant under the time reparametrization. Nonetheless some authors indicated
a way to overcome this difficulty by defining these indices in a generally
covariant way (for example, see Refs.~\refcite{IM1, IM2}). However, it should
be emphasized that this problem turns out not to be so significant for the
essence of the phenomenon. In fact Lyapunov indices are not an appropriate
instrument in general relativity but this does not mean that one cannot
identify a chaotic regime because many other of its symptoms can be found.

The best demonstration of the truth of the last statement is the work of N.
Cornish and J. Levin \cite{CL1,CL2} published in 1997. This was the final
persuasive confirmation of the chaoticity of the Bianchi type IX dynamics. The
authors used the two-dimensional Chernoff-Barrow map and showed the existence
of an attractor of fractal dimensionality, i.e., the strange attractor (the
same attractor which had been indicated by O. Bogoiavlenskii and S. Novikov,
but the latter authors at that time had not yet qualified it as a fractal
set). N. Cornish and J. Levin also calculated the topological entropy,
constructed a symbolic dynamics (codification of the trajectories by the words
and phrases of an alphabet) and showed that the discrete map they used
represents a valid approximation for the actual continuous evolution.

\section{Cosmological billiard}

Another efficient way to describe the oscillations near a cosmological
singularity as well as their stochastic character comes from the construction
of the so-called cosmological billiard \cite{DHN}. From the previous sections
it follows that the oscillatory regime arises in the first instance due to the
non-zero values of the quantities $\lambda,\mu,\nu$ of Eq.~(\ref{OR3-1}),
namely to the $\left(  a\right)  \neq\left(  b\right)  \neq\left(  c\right)  $
components of the quantities $l_{(a)}^{\mu}l_{(b)}^{\nu}[l_{\mu,\nu}%
^{(c)}-l_{\nu,\mu}^{(c)}]$. Then to clarify the basic properties of this
regime it will be useful to study simplified cosmological models with the
metric (\ref{OR1}) for which all three parameters $\lambda,\mu,\nu$ are
non-zero constants and the scale factors $a^{2},b^{2},c^{2}$ are functions
depending only on time. Moreover all of the other six quantities $l_{(a)}%
^{\mu}l_{(b)}^{\nu}[l_{\mu,\nu}^{(c)}-l_{\nu,\mu}^{(c)}]$ (apart of
$\lambda,\mu,\nu$) we can set to zero since they do not play any notable role
in creating the oscillations. It is a nontrivial fact that such exact models
indeed exist. These are the so-called diagonal homogeneous cosmological models
of Bianchi types IX and VIII (diagonality means that the projections
$g_{\alpha\beta}l_{(a)}^{\alpha}l_{(b)}^{\alpha}$ of the metric tensor onto
the frame of the Kasner directional vectors form the diagonal matrix with
respect to the frame indices) and the constant parameters $l_{(a)}^{\mu
}l_{(b)}^{\nu}[l_{\mu,\nu}^{(c)}-l_{\nu,\mu}^{(c)}]$ are nothing else but the
structural constants of the corresponding isometry group. The oscillatory
cosmological regime had been discovered with the help of these models in Ref.~\refcite{BK1}.

If instead of the scale factors $a^{2},b^{2},c^{2}$ we introduce the new
variables $\beta^{\left(  a\right)  }$ in accordance with notations:%
\begin{equation}
a^{2}=e^{-2\beta^{\left(  1\right)  }},\text{ }b^{2}=e^{-2\beta^{\left(
2\right)  }},\text{ }c^{2}=e^{-2\beta^{\left(  3\right)  }},\label{OR3-5}%
\end{equation}
and instead of the synchronous time $t$ the new time variable $\tau$ by the
relation:
\begin{equation}
d\tau=-\left(  abc\right)  ^{-1}dt\text{ },\label{OR3-6}%
\end{equation}
then it is easy to show that the Einstein equations for the aforementioned
Bianchi models in vacuum follow from the Lagrangian:%
\begin{equation}
L=T-V,\label{OR3-7}%
\end{equation}
where the \textquotedblleft kinetic energy" $T$ and \textquotedblleft
potential energy" $V$ are%
\begin{equation}
T=\left(  \frac{d\beta^{\left(  1\right)  }}{d\tau}\right)  ^{2}+\left(
\frac{d\beta^{\left(  2\right)  }}{d\tau}\right)  ^{2}+\left(  \frac
{d\beta^{\left(  3\right)  }}{d\tau}\right)  ^{2}-\left(  \frac{d\beta
^{\left(  1\right)  }}{d\tau}+\frac{d\beta^{\left(  2\right)  }}{d\tau}%
+\frac{d\beta^{\left(  3\right)  }}{d\tau}\right)  ^{2},\label{OR3-8}%
\end{equation}%
\begin{align}
V  & =\frac{1}{2}\lambda^{2}e^{-4\beta^{\left(  1\right)  }}+\frac{1}{2}%
\mu^{2}e^{-4\beta^{\left(  2\right)  }}+\frac{1}{2}\nu^{2}e^{-4\beta^{\left(
3\right)  }}\label{OR3-9}\\
& -\lambda\mu e^{-2\beta^{\left(  1\right)  }-2\beta^{\left(  2\right)  }%
}-\lambda\nu e^{-2\beta^{\left(  1\right)  }-2\beta^{\left(  3\right)  }}%
-\mu\nu e^{-2\beta^{\left(  2\right)  }-2\beta^{\left(  3\right)  }}.\nonumber
\end{align}
Here $\beta^{\left(  a\right)  }$ and $d\beta^{\left(  a\right)  }/d\tau$ are
considered to be generalized coordinates and generalized velocities. Apart
from the three differential equations of second order for $\beta_{\text{ }%
}^{\left(  a\right)  }$ which follow from this Lagrangian and which reproduce
the diagonal components of the equations $R_{\alpha\beta}=0$ (the non-diagonal
components of $R_{\alpha\beta}$ are identically zero), there is the well-known
additional constraint%
\begin{equation}
T+V=0,\label{OR3-10}%
\end{equation}
which is equivalent to the equation $R_{00}=0$ (the equations $R_{0\alpha}=0 $
for these models are satisfied identically). These are the exact Einstein
equations for the diagonal models of IX and VIII types in empty space. It is
easy to prove that in the asymptotic vicinity of the singularity, only the
first three terms in the potential (\ref{OR3-9}) survive, namely that in the
main approximation we have:%
\begin{equation}
V_{\left(  t\rightarrow0\right)  }=\frac{1}{2}\left(  \lambda^{2}%
e^{-4\beta^{\left(  1\right)  }}+\mu^{2}e^{-4\beta^{\left(  2\right)  }}%
+\nu^{2}e^{-4\beta^{\left(  3\right)  }}\right)  .\label{Vmain}%
\end{equation}
This is because the products $a^{2}b^{2},a^{2}c^{2},b^{2}c^{2}$ in the limit
$t\rightarrow0$ behave as $t^{2p_{\left(  a\right)  }+2p_{\left(  b\right)  }%
}$ for $a\neq b$ and they monotonically tend to zero due to the positiveness
of all three sums of Kasner exponents $p_{\left(  a\right)  }+p_{\left(
b\right)  } $ for $a\neq b$. The quantities $\lambda,\mu,\nu$ entering the
potential are the arbitrary constants but without loss of generality one can
take $\lambda=\mu=\nu=1$ for type IX and $\lambda=\mu=1,$ $\nu=-1$ for type
VIII. However, since the potential in the main approximation contains these
constants only in squares there is no difference in the asymptotic dynamics of
the models of types VIII and IX.

Owing to this Lagrangian representation the asymptotic evolution of the scale
factors can be described as a motion of a particle in the external potential
$V$. The earlier mentioned ``dangerous" space gradients which are responsible
for the instability of the Kasner dynamics are represented here by constants
$\lambda,\mu,\nu$ and the terms containing them create the reflecting
potential's walls responsible for the oscillatory regime. For the case of the
diagonal homogeneous cosmological model of the Bianchi IX type an effective
potential was introduced by C. Misner and D. Chitre \cite{Misn,Chitre}.
However, there is an essential difference between the Misner-Chitre approach
and the contemporary point of view of this mechanical analogy. The subtlety is
that the kinetic energy (\ref{OR3-8}) is an indefinite quadratic form with
respect to the velocities. To construct a system similar to a standard
mechanical one, it is necessary to find a way to single out those degrees of
freedom corresponding to a positive-definite kinetic energy form and then
include all other variables in the effective external potential. This method
was used in the Misner-Chitre construction. However, such an effective
external potential and its reflecting walls unavoidably evolve in time which
produces undesirable complications (especially for the succeeding analysis of
the general inhomogeneous gravitational field and its multidimensional generalization).

The present-day point of view was born around fifteen years ago in the work of
T. Damour, M. Henneaux and H. Nicolai (DHN)\cite{DHN} and according to them
the kinetic energy term (\ref{OR3-8}) should be taken exactly as it stands,
that is one should deal with the mechanics of the indefinite form of the
kinetic energy in the space of the Minkowski signature covered by the
coordinates $\beta^{\left(  a\right)  }$. In this case in the vicinity of the
singularity the potential and its reflecting walls remain stationary which
makes the analysis of their influence and their geometrical structure much
more transparent. Surprisingly enough this generalization of the mechanical
analogy turned out to be not only a technical facilitation but it led directly
to the discovery of the intriguing connection between the asymptotic
oscillatory regime and infinite-dimensional Kac-Moody algebras. This is the
consequence of the known mathematical fact that the root vectors of these
algebras are living in a space of Minkowski signature (unlike the root vectors
of the classical finite-dimensional Lie algebras which have to be built in
Euclidian space) and of that DHN observation that near the singularity all
dihedral angles between the reflecting walls acquire exactly those
distinguished values necessary for the vectors orthogonal to the walls to form
the set of simple roots of a Kac-Moody algebra.

It easy to see that the kinetic term (\ref{OR3-8}) corresponds to the motion
of a particle in the flat 3-dimensional $\beta$-space (with coordinates
$\beta^{\left(  a\right)  }$) of Minkowski signature\footnote{The standard
Minkowski space line element results from (\ref{OR3-11}) after the linear
coordinate transformation $\beta^{\left(  1\right)  }=\frac{1}{\sqrt{6}%
}\left(  \acute{\beta}^{\left(  1\right)  }-\acute{\beta}^{\left(  2\right)
}-\sqrt{3}\acute{\beta}^{\left(  3\right)  }\right)  ,$ $\beta^{\left(
2\right)  }=\frac{1}{\sqrt{6}}\left(  \acute{\beta}^{\left(  1\right)
}-\acute{\beta}^{\left(  2\right)  }+\sqrt{3}\acute{\beta}^{\left(  3\right)
}\right)  ,$ $\beta^{\left(  3\right)  }=\frac{1}{\sqrt{6}}\left(
\acute{\beta}^{\left(  1\right)  }+2\acute{\beta}^{\left(  2\right)  }\right)
.$ In coordinates $\acute{\beta}^{\left(  a\right)  }$ the interval becomes
$G_{\left(  a\right)  \left(  b\right)  }d\beta_{\text{ }}^{\left(  a\right)
}d\beta_{\text{ }}^{\left(  b\right)  }=-\left[  d\acute{\beta}^{\left(
1\right)  }\right]  ^{2}+\left[  d\acute{\beta}^{\left(  2\right)  }\right]
^{2}+\left[  d\acute{\beta}^{\left(  3\right)  }\right]  ^{2}.$} which has the
metric tensor $G_{\left(  a\right)  \left(  b\right)  }$ following from the
relation:%
\begin{equation}
G_{\left(  a\right)  \left(  b\right)  }d\beta_{\text{ }}^{\left(  a\right)
}d\beta_{\text{ }}^{\left(  b\right)  }=\sum_{\left(  a\right)  }\left[
d\beta_{\text{ }}^{\left(  a\right)  }\right]  ^{2}-\left[  \sum_{\left(
a\right)  }d\beta_{\text{ }}^{\left(  a\right)  }\right]  ^{2}.\label{OR3-11}%
\end{equation}

The constraint (\ref{OR3-10}) now can be written as%
\begin{equation}
G_{\left(  a\right)  \left(  b\right)  }\frac{\partial\beta^{\left(  a\right)
}}{\partial\tau}\frac{\partial\beta^{\left(  b\right)  }}{\partial\tau
}+V(\beta)=0.\label{OR3-12}%
\end{equation}
From this constraint and the fact that $V_{\left(  t\rightarrow0\right)  }>0$
follows that each trajectory $\beta^{\left(  a\right)  }(\tau)$ becomes
``time-like" with respect \ to the metric $G_{\left(  a\right)  \left(
b\right)  }$, i.e., near the singularity we have $G_{\left(  a\right)  \left(
b\right)  }\frac{\partial\beta^{\left(  a\right)  }}{\partial\tau}%
\frac{\partial\beta^{\left(  b\right)  }}{\partial\tau}<0.$ However, we
already know that in the limit $\tau\rightarrow\infty$ all scale factors tend
to zero, that is all three exponents\ $\beta^{\left(  a\right)  }(\tau)$ tend
to\ plus infinity, consequently each term in the potential $V$ tends to zero.
From this follows that trajectories $\beta^{\left(  a\right)  }(\tau)$ are
``time-like" only in the extreme vicinity of the potential walls
$\beta^{\left(  a\right)  }=0.$ Between the walls where $\beta^{\left(
a\right)  }$ are positive and very large the potential is exponentially small
and trajectories effectively become \textquotedblleft light-like", i.e.,
$G_{\left(  a\right)  \left(  b\right)  }\frac{\partial\beta^{\left(
a\right)  }}{\partial\tau}\frac{\partial\beta^{\left(  b\right)  }}%
{\partial\tau}=0. $ These periods of \textquotedblleft light-like" motion
between the walls exactly correspond to the Kasner epochs. It is easy to see
that the walls themselves are \textquotedblleft time-like". Indeed, the three
vectors $w_{A}$ of the $\beta$-space with frame components $w_{A\left(
a\right)  }$, that is $w_{A}=\left(  w_{A\left(  1\right)  },w_{A\left(
2\right)  },w_{A\left(  3\right)  }\right)  $ (here and below uppercase Latin
indices\ take values $1,2,3$\ and enumerate the walls), orthogonal to the
walls $\beta_{\text{ }}^{\left(  1\right)  }=0,$ $\beta_{\text{ }}^{\left(
2\right)  }=0,$ $\beta_{\text{ }}^{\left(  3\right)  }=0$ are $w_{1}=(1,0,0),$
$w_{2}=(0,1,0),$ $w_{3}=(0,0,1)$ respectively and all three have one and the
same positive norm $G^{\left(  a\right)  \left(  b\right)  }w_{A\left(
a\right)  }w_{A\left(  b\right)  }=1/2$ (where $G^{\left(  a\right)  \left(
b\right)  }$ is inverse to $G_{\left(  a\right)  \left(  b\right)  }$ and
there is no summation over $A$), that is they are \textquotedblleft
space-like" which means that the walls are \textquotedblleft time-like" and
collisions of a \textquotedblleft light-like" particle against such walls are
inescapable and interminable.

It is worth emphasizing that in this analysis we consider the potential walls
as (effectively) infinitely sharp and of infinite height. It is quite
remarkable that in the asymptotic vicinity of the singularity this becomes
really an acceptable picture. This is one of the essential points observed in
Ref.~\refcite{DHN}. Such asymptotic properties of the walls simplify further
the dynamics near the singularity and make transparent the reasons for the
chaotic nature of such oscillatory dynamics.

Since $G_{\left(  a\left(  b\right)  \right)  }\beta_{\text{ }}^{\left(
a\right)  }\beta_{\text{ }}^{\left(  b\right)  }=-2(\beta^{\left(  1\right)
}\beta^{\left(  2\right)  }+\beta^{\left(  1\right)  }\beta^{\left(  3\right)
}+\beta^{\left(  2\right)  }\beta^{\left(  3\right)  })$ and near the
singularity all $\beta_{\text{ }}^{\left(  a\right)  }$ are positive we have
$G_{\left(  a\left(  b\right)  \right)  }\beta_{\text{ }}^{\left(  a\right)
}\beta_{\text{ }}^{\left(  b\right)  }<0$. Then one can introduce instead of
$\beta_{\text{ }}^{\left(  a\right)  }$ the ``radial" coordinate $\rho>0$
($\rho\rightarrow\infty$ when $\tau\rightarrow\infty$) and ``angular"
coordinates $\gamma^{\left(  a\right)  }$ in the following way:%
\begin{equation}
\beta_{\text{ }}^{\left(  a\right)  }=\rho\gamma^{\left(  a\right)  },\text{
\ }G_{\left(  a)\left(  b\right)  \right)  }\gamma_{\text{ }}^{\left(
a\right)  }\gamma_{\text{ }}^{\left(  b\right)  }=-1.\label{Hom81}%
\end{equation}
Now we have:
\begin{equation}
G_{\left(  a\right)  \left(  b\right)  }\frac{\partial\beta^{\left(  a\right)
}}{\partial\tau}\frac{\partial\beta^{\left(  b\right)  }}{\partial\tau
}=-\left(  \frac{\partial\rho}{\partial\tau}\right)  ^{2}+\rho^{2}G_{\left(
a\right)  \left(  b\right)  }\frac{\partial\gamma^{\left(  a\right)  }%
}{\partial\tau}\frac{\partial\gamma^{\left(  b\right)  }}{\partial\tau
}.\label{Hom82}%
\end{equation}
The second condition in (\ref{Hom81}) picks out $\ $in $\gamma$-space the
two-dimensional Lobachevsky surface. Indeed, introducing two variables
$r,\phi:$
\begin{align}
\gamma^{\left(  1\right)  }  & =\frac{1}{\sqrt{6}}\cosh r-\frac{1}{\sqrt{6}%
}\left(  \sin\phi+\sqrt{3}\cos\phi\right)  \sinh r,\label{Hom83}\\
\gamma^{\left(  2\right)  }  & =\frac{1}{\sqrt{6}}\cosh r-\frac{1}{\sqrt{6}%
}\left(  \sin\phi-\sqrt{3}\cos\phi\right)  \sinh r,\nonumber\\
\gamma^{\left(  3\right)  }  & =\frac{1}{\sqrt{6}}\cosh r+\frac{2}{\sqrt{6}%
}\sin\phi\sinh r,\nonumber
\end{align}
it is easy to check that for any values of $r,\phi$ the condition $G_{\left(
a)\left(  b\right)  \right)  }\gamma_{\text{ }}^{\left(  a\right)  }%
\gamma_{\text{ }}^{\left(  b\right)  }=-1$ is satisfied identically and the
metric on any arbitrarily chosen hypersurface $\rho=const,$ imbedded into the
$\beta$-space, is:%
\begin{equation}
\left(  G_{\left(  a\right)  \left(  b\right)  }d\beta_{\text{ }}^{\left(
a\right)  }d\beta_{\text{ }}^{\left(  b\right)  }\right)  _{\rho=const}%
=\rho^{2}G_{\left(  a\right)  \left(  b\right)  }d\gamma_{\text{ }}^{\left(
a\right)  }d\gamma_{\text{ }}^{\left(  b\right)  }=\rho^{2}\left(
dr^{2}+\sinh^{2}rd\phi^{2}\right)  .\label{Hom84}%
\end{equation}

This 2-dimensional space covered by coordinates $r,\phi$ is the Lobachevsky
surface of constant negative curvature and each trajectory $\beta^{\left(
a\right)  }(\tau)$ has the radially projected trace on this surface. It can be
shown that the free Kasner flights\ in three-dimensional $\beta$-space between
the walls are projected into \textit{geodesics} on this two-dimensional surface.

The walls $\beta_{\text{ }}^{\left(  1\right)  }=0,$ $\beta_{\text{ }%
}^{\left(  2\right)  }=0,$ $\beta_{\text{ }}^{\left(  3\right)  }=0,$ as can
be seen from (\ref{Hom81}) and (\ref{Hom83}), are projected into the three
curves respectively:
\begin{equation}
\coth r=2\sin\left( \phi+\frac{\pi}{3}\right) ,\text{ } \coth r=2\sin\left(
\phi-\frac{\pi}{3}\right) ,\text{ } \coth r=-2\sin\phi,\label{Hom91}%
\end{equation}
forming a triangle on the Lobachevsky surface, a triangle which has all three
vertices at infinity and all its three angles equal to zero. The projection of
a trajectory $\beta^{\left(  a\right)  }(\tau)$ into this surface represents
the endless geodesic motion of a \textquotedblleft\textit{billiard ball}"
between the cushions (\ref{Hom91}) of this triangular \textquotedblleft%
\textit{billiard table}".\emph{\ }It can be shown that reflections against the
cushions of this triangular billiard table are of mirror type and the
two-dimensional volume of this billiard is finite in spite of its non-compactness.

Then the facts are: (i) the billiard table represents a finite part of the
2-dimensional space of constant negative curvature, (ii) the volume of the
billiard is finite, (iii) the 2-dimensional trajectories of the billiard ball
between reflections against the cushions are geodesics, (iv) the reflections
are of the mirror type, that is the angles of incidence and reflections are
the same. It is well known that the motion under these four conditions is
strongly chaotic (including mixing). This conclusion follows mainly from the
works of E. Hopf \cite{H}, D. Anosov \cite{An} and Ya. Sinai \cite{S1,S2} and
this is in agreement with the chaotic character of the oscillatory
cosmological regime derived solely on the base of the evolution of the Kasner
exponents as explained in the second section of the present paper.

\section{First manifestation of the hidden Kac-Moody algebra}

The preceding results show that in spite of complicated structure of the
gravitational field its asymptotics near cosmological singularity is quite
understandable. The billiard table we described in the previous section
represents the generalization to the covering space of the simplest Coxeter
crystallographic simplex, that is the equilateral triangle in 2-dimensional
$\gamma$-space (Lobachevsky plane) having its three vertices at infinity with
all angles between the faces equal to zero. This is a very special geometric
construction which (when combined with the specular laws of reflections
against the faces) indicates the nontrivial symmetry hidden in the space-time
near the cosmological singularity.

The mathematical description of the symmetry we are talking about can be
achieved in the following way. Consider the trajectories $\beta_{\text{ }%
}^{\left(  a\right)  }(\tau)$ of a particle moving between the walls
$\beta_{\text{ }}^{\left(  a\right)  }=0$ in the original 3-dimensional
$\beta$-space with coordinates $\beta_{\text{ }}^{\left(  a\right)  }$ and
metric $G_{\left(  a\right)  \left(  b\right)  }$ of Eq.~(\ref{OR3-11}). These
trajectories are null straight lines with respect to the metric $G_{\left(
a\right)  \left(  b\right)  }.$ We mentioned already in the previous section
that three vectors $w_{A}$ with frame components $w_{A\left(  a\right)  }$,
that is $w_{A}=\left(  w_{A\left(  1\right)  },w_{A\left(  2\right)
},w_{A\left(  3\right)  }\right)  ,$ orthogonal to the walls $\beta_{\text{ }%
}^{\left(  1\right)  }=0,$ $\beta_{\text{ }}^{\left(  2\right)  }=0,$
$\beta_{\text{ }}^{\left(  3\right)  }=0$ are $w_{1}=(1,0,0),$ $w_{2}%
=(0,1,0),$ $w_{3}=(0,0,1)$ respectively. We can imagine these vectors as
arrows starting at the origin of the $\beta$-space. They have fixed finite
norm and one can arrange the scalar products $(w_{A}\bullet w_{B})=G^{\left(
a\right)  \left(  b\right)  }w_{A\left(  a\right)  }w_{B\left(  b\right)  }$
in the form of the matrix:%
\begin{equation}
A_{AB}=2\frac{(w_{A}\bullet w_{B})}{(w_{A}\bullet w_{A})}\text{ \ \ (no
summation in }A\text{).}\label{Kac1}%
\end{equation}
The calculation gives:%
\begin{equation}
A_{AB}=\left(
\begin{array}
[c]{ccc}%
2 & -2 & -2\\
-2 & 2 & -2\\
-2 & -2 & 2
\end{array}
\right) \label{Kac2}%
\end{equation}
The point is that $A_{AB}$ is the Cartan matrix of indefinite type, i.e., with
one negative principal value. Any Cartan matrix can be associated with some
Lie algebra and particular matrix (\ref{Kac2}) corresponds to one of the
Lorenzian hyperbolic Kac-Moody algebras of rank 3 (it has number 7 in the list
of 19 possible algebras of this type provided in Ref.~\refcite{Sac}).

As follows from the transition law (\ref{OR4})--(\ref{OR5}) between two Kasner
epochs and the notations (\ref{OR3-5})--(\ref{OR3-6}), the particle's velocity
$v^{\left(  a\right)  }=d\beta^{\left(  a\right)  }/d\tau$ before and after
reflection from the wall $\beta_{\text{ }}^{\left(  1\right)  }=0$ have the
following components: $(v^{\left(  a\right)  })_{before}=\left(  qp_{\left(
1\right)  },\text{ }qp_{\left(  2\right)  },\text{ }qp_{\left(  3\right)
}\right)  $ and $(v^{\left(  a\right)  })_{after}=\left(  -qp_{\left(
1\right)  },\text{ }qp_{\left(  2\right)  }+2qp_{\left(  1\right)  },\text{
}qp_{\left(  3\right)  }+2qp_{\left(  1\right)  }\right)  $ where
$q=a_{0}b_{0}c_{0}$. Using the vector $w_{1}=(1,0,0)$ orthogonal to this wall
and his contravariant components $G^{\left(  a\right)  \left(  b\right)
}w_{1\left(  b\right)  }=G^{\left(  a\right)  \left(  1\right)  }=\left(
1/2,\text{ }-1/2,\text{ }-1/2\right)  \ $the change of the velocity is
$(v^{\left(  a\right)  })_{after}=(v^{\left(  a\right)  })_{before}%
-2(v^{\left(  b\right)  })_{before}w_{1\left(  b\right)  }(w_{1}\bullet
w_{1})^{-1}G^{\left(  a\right)  \left(  c\right)  }w_{1\left(  c\right)  }.$
The relations of the same types are valid for the velocity change during the
reflections from two other walls (in the last formula we should just replace
the vector $w_{1}$ by $w_{2}$ or $w_{3}$). Then the law governing reflections
against each wall ($A=1,2,3$) can be unified in the following relations:
\begin{equation}
(v^{\left(  a\right)  })_{after}=(v^{\left(  a\right)  })_{before}%
-2\frac{(v^{\left(  b\right)  })_{before}w_{A\left(  b\right)  }}%
{(w_{A}\bullet w_{A})}G^{\left(  a\right)  \left(  c\right)  }w_{A\left(
c\right)  }\text{ },\text{(no summation in }A\text{).}\label{Kac3}%
\end{equation}
This transformation represents the specular reflection of a particle by the
wall orthogonal to the vector $w_{A}$ and it is nothing else but the standard
form of Weyl reflections. Now it is clear that one can formally identify the
three vectors $w_{A}$ with the simple roots of the root system of the
corresponding Kac-Moody algebra, the walls $\beta_{\text{ }}^{\left(
a\right)  }=0$ with the Weyl hyperplanes orthogonal to the simple roots, the
reflections (\ref{Kac3}) with the elements of the Weyl group of the root
system and the region of $\beta$-space bounded by the walls (in which a
particle oscillates) with the fundamental Weyl chamber. All these results can
be found in Ref.~\refcite{DHN} and in the exhaustive review \refcite{HPS}
which has been written especially for applications of the generalized Lie
algebras to the theory of the cosmological singularity in general relativity
and to string theories.

The appearance of Lie algebras associated with properties of the cosmological
singularity gave birth to the so-called Hidden Symmetry Conjecture which
assumes that the manifestation of these algebras means that the corresponding
infinite-dimensional Lie symmetry group might somehow be hidden in the system.
This conjecture proposes that such symmetry might be inherent in the exact
theories and not only for their limits in the vicinity of the cosmological
singularity. If so the limiting structure near the singularity should be
considered just as an auxiliary instrument by means of which this symmetry is
coming to light. In fact from the DHN approach it follows that the appearance
of infinite-dimensional Kac-Moody algebras in connection with the oscillatory
character of the cosmological singularity turns out to be a much more general
phenomenon than what we observed here using just the diagonal Bianchi type IX
and Bianchi type VIII homogeneous models. The crucial point is that majority
of the physically reasonable cases, including supergravities emerging in the
low energy limit from all types of superstring models, have the BKL
oscillatory cosmological singularity in the heart of which lies that or
another type of the hyperbolic Kac-Moody algebra. The appearance of these
algebras might signify the nontrivial huge symmetry hidden in the asymptotic
structure of space-time near the cosmological singularity, a symmetry which
neverltheless coexists with chaoticity. As of now we have no understanding
where and how exactly this symmetry would act (it could be as a continuous
infinite-dimensional symmetry group of the exact Lagrangian permitting the
transformation of the given solutions of the equations of motion to the new
solutions). If true the hidden symmetry conjecture could create new
inestimable perspectives for the development of relativistic gravity and
superstring theories.

\section{Rotation of Kasner axis and frozen directions}

In the two previous sections we focused attention on the behavior of the scale
factors with respect to stationary Kasner axes. Although this was very useful
for understanding some basic properties of the oscillatory regime such an
approach is insufficient for the description of the most general behavior of
the metric tensor near a cosmological singularity. The point is that in the
course of each Kasner epoch these axes remain stationary but from epoch to
epoch (mainly during the short transition period between the epochs) they
change their directions. We call this evolution the \textit{rotation of the
Kasner axes }and this effect has been described in Ref.~\refcite{BKL1}. We
emphasize again that the laws governing the evolution of the numerical values
of the Kasner exponents (described in our second section) do not depend on how
Kasner axes change their directions from epoch to epoch and that this
invariance makes the behavior of Kasner exponents a particularly important
aspect of the oscillatory regime. Nevertheless, for the complete description
of the asymptotics near the singularity one need to know also the evolution of
the directional vectors $l_{\alpha}^{\left(  a\right)  }$ of the Kasner axes
and influence of this evolution on the structure of cosmological billiard.

In general in order to take into account the rotation of the frame we can
represent its directional vectors by linear combinations of their initial
values with time-dependent coefficients (for example, of the vectors directed
along the Kasner axes of the initial epoch). Then instead of (\ref{OR1}) we
have to write
\begin{equation}
g_{\alpha\beta}=\Gamma_{\left(  a\right)  \left(  b\right)  }\left[
A_{\left(  c\right)  }^{\left(  a\right)  }l_{\alpha}^{\left(  c\right)
}\right]  \left[  A_{\left(  d\right)  }^{\left(  b\right)  }l_{\beta
}^{\left(  d\right)  }\right]  \text{ },\text{ \ }\Gamma_{\left(  a\right)
\left(  b\right)  }= \mathrm{diag}\left[  \Gamma_{\left(  1\right)  }%
,\Gamma_{\left(  2\right)  },\Gamma_{\left(  3\right)  }\right]  ),\label{1}%
\end{equation}
where $l_{\alpha}^{\left(  a\right)  }=l_{\alpha}^{\left(  a\right)  }%
(x^{1},x^{2},x^{3})$ is some initial frame and the time-dependent matrix
$A_{\left(  c\right)  }^{\left(  a\right)  }(t,x^{1},x^{2},x^{3})$ is
responsible for the rotation effect we are interested in. The components of
the diagonal matrix $\Gamma_{\left(  a\right)  \left(  b\right)  }%
(t,x^{1},x^{2},x^{3})$ we continue to call the scale factors but now with
respect to the rotating frame $A_{\left(  b\right)  }^{\left(  a\right)
}l_{\alpha}^{\left(  b\right)  }$ and we designate these new scale factors by
the new letters $\Gamma_{\left(  a\right)  }$ to emphasize that they are
different from $a^{2},b^{2},c^{2}$ in (\ref{OR1}). It is convenient to single
out these new scale factors $\Gamma_{\left(  a\right)  }$ explicitly, that is,
not absorb them into the rotation matrix $A_{\left(  b\right)  }^{\left(
a\right)  }.$ With such a decomposition we have only three physical dynamical
degrees of freedom in this matrix, and another three degrees we include in the
scale factors $\Gamma_{\left(  a\right)  }.$

The axes of a rotating Kasner frame can be defined in an rigorous mathematical
way by the condition that both the metric tensor $g_{\alpha\beta}$ and the
second fundamental form $\dot{g}_{\alpha\beta}$ when projected onto this frame
are diagonal (i.e., are an orthogonal frame of eigenvectors of the extrinsic
curvature). It is always possible to find such a global Kasner frame using the
gauge freedom to rotate the orthonormal triad $L_{\alpha}^{\left(  a\right)
}=\gamma_{\left(  b\right)  }^{\left(  a\right)  }A_{\left(  c\right)
}^{\left(  b\right)  }l_{\alpha}^{\left(  c\right)  }$ where $\gamma_{\left(
b\right)  }^{\left(  a\right)  }=\mathrm{diag}(\sqrt{\Gamma_{\left(  1\right)
}},\sqrt{\Gamma_{\left(  2\right)  }},\sqrt{\Gamma_{\left(  3\right)  }}).$

However, the effect of the rotation of the Kasner axes make their use
inconvenient for the analytical description of the asymptotic regime near the
singularity because the laws governing these rotations are complicated enough
and the rotations of Kasner axes never stop all the way to the singularity.
Fortunately, it turns out that another set of axes exists, rotations of which
do stop in the limit $t\rightarrow0$ and projection of the metric tensor (but
not of the second form $\dot{g}_{\alpha\beta}$) onto such an asymptotically
\textquotedblleft frozen" 3-dimensional frame still is a diagonal matrix. It
turns out that the components of this diagonal matrix (that is the scale
factors with respect to these new frozen axes) have no limit since their
behavior again can be described by an infinite sequence of alternating
Kasner-like epochs with the same transformation laws of the numerical values
of the power law exponents as before. To find such \textit{frozen axes} is an
efficient way to reduce the description of the asymptotic evolution of the six
components of the metric tensor to the three oscillating degrees of freedom.

Due to the gauge freedom of frame rotations the choice of the matrix
$A_{\left(  b\right)  }^{\left(  a\right)  }$ in the metric (\ref{1}) is not
unique. One convenient choice follows from the fact that any symmetric matrix
can be diagonalized by an orthogonal matrix, that is $A_{\left(  b\right)
}^{\left(  a\right)  }$ in the metric tensor (\ref{1}) can be chosen to be an
orthogonal matrix $O_{\left(  b\right)  }^{\left(  a\right)  }$ without loss
of generality. In this case the factor $\Gamma_{\left(  a\right)  \left(
b\right)  }O_{\left(  c\right)  }^{\left(  a\right)  }O_{\left(  d\right)
}^{\left(  b\right)  }$ in metric (\ref{1}) contains six dynamical degrees of
freedom (three Euler angles in $O_{\left(  b\right)  }^{\left(  a\right)  }$
and three scale factors in the diagonal matrix $\Gamma_{\left(  a\right)
\left(  b\right)  }$) as is the case for any 3-dimensional metric tensor. The
auxiliary vectors $l_{\alpha}^{\left(  a\right)  }(x^{1},x^{2},x^{3})$ can be
taken in any desirable form which can be chosen on the basis of technical
convenience. If we wish to generalize the diagonal Bianchi type IX and VIII
models to their most general non-diagonal and inhomogeneous versions it is
natural to fix vectors $l_{\alpha}^{\left(  a\right)  }$ exactly in the same
way as before, that is to set all three parameters $\lambda,\mu,\nu$
(\ref{OR3-1}) to be non-zero constants and set all the other six quantities
$l_{(a)}^{\mu}l_{(b)}^{\nu}[l_{\mu,\nu}^{(c)}-l_{\nu,\mu}^{(c)}]$ (apart from
$\lambda,\mu,\nu$) to zero. If in addition to this we restrict the matrix
$\Gamma_{\left(  a\right)  \left(  b\right)  }O_{\left(  c\right)  }^{\left(
a\right)  }O_{\left(  d\right)  }^{\left(  b\right)  }$ to be dependent only
on time we obtain the most general homogeneous models of Bianchi type IX and
VIII. The freezing effect and dynamics of the new scale factors $\Gamma
_{\left(  a\right)  \left(  b\right)  }$ defined with respect to the new
rotating axes $O_{\left(  b\right)  }^{\left(  a\right)  }l_{\alpha}^{\left(
b\right)  }$ (we emphasize again that they are different from the rotating
Kasner axes) we will demonstrate, namely in the framework of these homogeneous
generalizations and for definiteness only for the Bianchi type IX model.

\section{The general Bianchi type IX model}

In accordance with the previous discussion the metric for the most general
Bianchi type IX cosmological model can be taken in the form:%
\begin{equation}
g_{\alpha\beta}=\Gamma_{\left(  a\right)  \left(  b\right)  }O_{\left(
c\right)  }^{\left(  a\right)  }O_{\left(  d\right)  }^{\left(  b\right)
}l_{\alpha}^{\left(  c\right)  }l_{\beta}^{\left(  d\right)  }\text{ },\text{
\ }\Gamma_{\left(  a\right)  \left(  b\right)  }=\mathrm{diag}\left[
\Gamma_{\left(  1\right)  },\Gamma_{\left(  2\right)  },\Gamma_{\left(
3\right)  }\right]  ),\label{2}%
\end{equation}
where the vectors $l_{\alpha}^{\left(  a\right)  }$ satisfy Eqs.~(\ref{OR3-1})
with $\lambda,\mu,\nu=const$\ and all other quantities $l_{(a)}^{\mu}%
l_{(b)}^{\nu}[l_{\mu,\nu}^{(c)}-l_{\nu,\mu}^{(c)}]$ apart from $\lambda
,\mu,\nu$ are zero. In this section to simplify the notation we set
$\lambda=\mu=\nu=1.$ The scale factors $\Gamma_{\left(  1\right)  }%
,\Gamma_{\left(  2\right)  },\Gamma_{\left(  3\right)  }$ and the orthogonal
matrix $O_{\left(  b\right)  }^{\left(  a\right)  }$ depend only on time and
we parametrize this matrix in the standard way by the Euler angles
$\psi,\theta,{\large \varphi}:$%
\begin{equation}
\text{$\left(
\begin{array}
[c]{ccc}%
\cos{\tiny \psi}\cos{\tiny \varphi-}\sin{\tiny \psi}\cos{\tiny \theta}%
\sin{\tiny \varphi} & \cos{\tiny \psi}\sin{\tiny \varphi+}\sin{\tiny \psi}%
\cos{\tiny \theta}\cos{\tiny \varphi} & \sin{\tiny \psi}\sin{\tiny \theta}\\
{\tiny -}\sin{\tiny \psi}\cos{\tiny \varphi-}\cos{\tiny \psi}\cos
{\tiny \theta}\sin{\tiny \varphi} & {\tiny -}\sin{\tiny \psi}\sin
{\tiny \varphi+}\cos{\tiny \psi}\cos{\tiny \theta}\cos{\tiny \varphi} &
\cos{\tiny \psi}\sin{\tiny \theta}\\
\sin{\tiny \theta}\sin{\tiny \varphi} & {\tiny -}\sin{\tiny \theta}%
\cos{\tiny \varphi} & \cos{\tiny \theta}%
\end{array}
\right)  $}{\small .}\label{IX-9}%
\end{equation}

Here we follow the rule: an upper index in $O_{\left(  b\right)  }^{\left(
a\right)  }$ enumerates the rows and lower index corresponds to columns.\ As
in case of diagonal models it is convenient to introduce instead of the
synchronous time $t$ the new variable $\tau$ by the relation
\begin{equation}
dt=-\left[  \det\left(  \Gamma_{\left(  a\right)  \left(  b\right)
}O_{\left(  c\right)  }^{\left(  a\right)  }O_{\left(  d\right)  }^{\left(
b\right)  }\right)  \right]  ^{\frac{1}{2}}d\tau=-\left[  \det\left(
\Gamma_{\left(  a\right)  \left(  b\right)  }\right)  \right]  ^{\frac{1}{2}%
}d\tau=-\left[  \Gamma_{\left(  1\right)  }\Gamma_{\left(  2\right)  }%
\Gamma_{\left(  3\right)  }\right]  ^{\frac{1}{2}}d\tau.\label{IX-9-A}%
\end{equation}
Now we define the \textquotedblleft angular velocities" $\Omega_{\left(
a\right)  }$ \
\begin{align}
\Omega_{\left(  1\right)  }  & =\frac{d\varphi}{d\tau}\sin\psi\sin\theta
+\frac{d\theta}{d\tau}\cos\psi,\label{IX-11}\\
\Omega_{\left(  2\right)  }  & =\frac{d\varphi}{d\tau}\cos\psi\sin\theta
-\frac{d\theta}{d\tau}\sin\psi,\nonumber\\
\Omega_{\left(  3\right)  }  & =\frac{d\varphi}{d\tau}\cos\theta+\frac{d\psi
}{d\tau},\nonumber
\end{align}
and three quantities $I_{\left(  1\right)  },I_{\left(  2\right)  },I_{\left(
3\right)  }$ which we call \textquotedblleft moments of inertia":%
\begin{equation}
I_{\left(  1\right)  }=\frac{\left(  \Gamma_{(2)}-\Gamma_{(3)}\right)  ^{2}%
}{\Gamma_{(2)}\Gamma_{(3)}},\text{ }I_{\left(  2\right)  }=\frac{\left(
\Gamma_{(1)}-\Gamma_{(3)}\right)  ^{2}}{\Gamma_{(1)}\Gamma_{(3)}},\text{
}I_{\left(  3\right)  }=\frac{\left(  \Gamma_{(1)}-\Gamma_{(2)}\right)  ^{2}%
}{\Gamma_{(1)}\Gamma_{(2)}}.\label{IX-12}%
\end{equation}
With this notation the six Einstein equations $R_{\alpha\beta}=0$
\footnote{The general Bianchi type IX model considered here , if understood as
the exact solution of the Einstein equations, can exist only in the space
filled with matter, otherwise the equations $R_{0\alpha}=0$ would lead to the
elimination of the rotational angles $\psi,\theta,\varphi$. However, in the
asymptotic vicinity of the singularity, which is the region of our interest,
the presence of the standard forms of matter (we will discuss some exceptional
cases in the section \textquotedblleft The influence of matter") has no
decisive importance in the $00$ and $\alpha\beta$ components of the Einstein
equations and in these equations in the main approximation we can neglect the
energy-momentum tensor. In the $0\alpha$ components the right hand side of the
gravitational equations is of the same order as the left hand side even in the
main approximation but we need not to consider these equations because they
put only some restriction on the parameters of the solution without direct
influence on the character of its dynamics. The inhomogeneous generalization
of the Bianchi type IX model considered here exists already in vacuum since
inhomogeneity imitates the presence of matter. An analysis of the exact (not
only near singularity) general Bianchi type IX model for the case of a perfect
fluid can be found in Ref.~\refcite{Bog}.} give the Euler equations for a
freely rotating (but not rigid) asymmetric top (from which follows the reason
for using the terminology \textquotedblleft moments of inertia"):%

\begin{align}
\frac{d}{d\tau}\left(  I_{\left(  1\right)  }\Omega_{\left(  1\right)
}\right)  +\left(  I_{\left(  3\right)  }-I_{\left(  2\right)  }\right)
\Omega_{\left(  2\right)  }\Omega_{\left(  3\right)  }  & =0,\label{IX-15}\\
\frac{d}{d\tau}\left(  I_{\left(  2\right)  }\Omega_{\left(  2\right)
}\right)  +\left(  I_{\left(  1\right)  }-I_{\left(  3\right)  }\right)
\Omega_{\left(  1\right)  }\Omega_{\left(  3\right)  }  & =0,\nonumber\\
\frac{d}{d\tau}\left(  I_{\left(  3\right)  }\Omega_{\left(  3\right)
}\right)  +\left(  I_{\left(  2\right)  }-I_{\left(  1\right)  }\right)
\Omega_{\left(  1\right)  }\Omega_{\left(  2\right)  }  & =0,\nonumber
\end{align}
and three equations governing the evolution of the scale factors (and
consequently of the ``moments of inertia"):
\begin{align}
& \frac{d^{2}\ln\Gamma_{(1)}}{d\tau^{2}}+\left(  \Gamma_{(1)}\right)
^{2}-\left(  \Gamma_{(2)}-\Gamma_{(3)}\right)  ^{2}\label{IX-17}\\
& \qquad-\frac{\Gamma_{(1)}\Gamma_{(2)}\left(  \Gamma_{(1)}+\Gamma
_{(2)}\right)  }{\left(  \Gamma_{(1)}-\Gamma_{(2)}\right)  ^{3}}\left(
I_{\left(  3\right)  }\Omega_{\left(  3\right)  }\right)  ^{2}-\frac
{\Gamma_{(1)}\Gamma_{(3)}\left(  \Gamma_{(1)}+\Gamma_{(3)}\right)  }{\left(
\Gamma_{(1)}-\Gamma_{(3)}\right)  ^{3}}\left(  I_{\left(  2\right)  }%
\Omega_{\left(  2\right)  }\right)  ^{2}=0,\nonumber\\
&  \frac{d^{2}\ln\Gamma_{(2)}}{d\tau^{2}}+\left(  \Gamma_{(2)}\right)
^{2}-\left(  \Gamma_{(1)}-\Gamma_{(3)}\right)  ^{2}\label{IX-18}\\
& \qquad+\frac{\Gamma_{(1)}\Gamma_{(2)}\left(  \Gamma_{(1)}+\Gamma
_{(2)}\right)  }{\left(  \Gamma_{(1)}-\Gamma_{(2)}\right)  ^{3}}\left(
I_{\left(  3\right)  }\Omega_{\left(  3\right)  }\right)  ^{2}-\frac
{\Gamma_{(2)}\Gamma_{(3)}\left(  \Gamma_{(2)}+\Gamma_{(3)}\right)  }{\left(
\Gamma_{(2)}-\Gamma_{(3)}\right)  ^{3}}\left(  I_{\left(  1\right)  }%
\Omega_{\left(  1\right)  }\right)  ^{2}=0,\nonumber\\
&  \frac{d^{2}\ln\Gamma_{(3)}}{d\tau^{2}}+\left(  \Gamma_{(3)}\right)
^{2}-\left(  \Gamma_{(1)}-\Gamma_{(2)}\right)  ^{2}\label{IX-19}\\
& \qquad+\frac{\Gamma_{(1)}\Gamma_{(3)}\left(  \Gamma_{(1)}+\Gamma
_{(3)}\right)  }{\left(  \Gamma_{(1)}-\Gamma_{(3)}\right)  ^{3}}\left(
I_{\left(  2\right)  }\Omega_{\left(  2\right)  }\right)  ^{2}+\frac
{\Gamma_{(2)}\Gamma_{(3)}\left(  \Gamma_{(2)}+\Gamma_{(3)}\right)  }{\left(
\Gamma_{(2)}-\Gamma_{(3)}\right)  ^{3}}\left(  I_{\left(  1\right)  }%
\Omega_{\left(  1\right)  }\right)  ^{2}=0.\nonumber
\end{align}

Finally, the equation $R_{00}=0$ (taking into account the previous equations
$R_{\alpha\beta}=0$) can be reduced to the form:%
\begin{align}
& \frac{1}{4}\left[  \left(  \frac{d\ln\Gamma_{(1)}}{d\tau}\right)
^{2}+\left(  \frac{d\ln\Gamma_{(2)}}{d\tau}\right)  ^{2}+\left(  \frac
{d\ln\Gamma_{(3)}}{d\tau}\right)  ^{2}\right] \label{IX-20}\\
& \qquad-\frac{1}{4}\left[  \frac{d\ln\Gamma_{(1)}}{d\tau}+\frac{d\ln
\Gamma_{(2)}}{d\tau}+\frac{d\ln\Gamma_{(3)}}{d\tau}\right]  ^{2}\nonumber\\
& \qquad+\frac{1}{2}I_{\left(  1\right)  }\left(  \Omega_{\left(  1\right)
}\right)  ^{2}+\frac{1}{2}I_{\left(  2\right)  }\left(  \Omega_{\left(
2\right)  }\right)  ^{2}+\frac{1}{2}I_{\left(  3\right)  }\left(
\Omega_{\left(  3\right)  }\right)  ^{2}\nonumber\\
& \qquad\frac{1}{2}\left[  \left(  \Gamma_{(1)}\right)  ^{2}+\left(
\Gamma_{(2)}\right)  ^{2}+\left(  \Gamma_{(3)}\right)  ^{2}-2\Gamma
_{(1)}\Gamma_{(2)}-2\Gamma_{(1)}\Gamma_{(3)}-2\Gamma_{(2)}\Gamma_{(3)}\right]
=0\nonumber
\end{align}

The last equation (\ref{IX-20}) represents the most important instrument for
the qualitative analyses of the dynamics of the model under consideration
because it provides the Lagrangian for the equations of motion (\ref{IX-15}%
)--(\ref{IX-19}). In a sense the constraint (\ref{IX-20}) (as usual in general
relativity) can be treated as the zero \textquotedblleft energy" condition if
the sum of all terms in the left hand side of (\ref{IX-20}) containing the
time derivatives we identify with the \textquotedblleft kinetic energy" and
the sum of all the remaining terms with the ``potential energy". In this way
we conclude that the Lagrangian, that is the difference between the
\textquotedblleft kinetic energy" and the \textquotedblleft potential
energy",$\mathcal{\ }$should be:%
\begin{align}
\mathcal{L}\text{ }  & \mathcal{=}\frac{1}{4}\left[  \left(  \frac{d\ln
\Gamma_{(1)}}{d\tau}\right)  ^{2}+\left(  \frac{d\ln\Gamma_{(2)}}{d\tau
}\right)  ^{2}+\left(  \frac{d\ln\Gamma_{(3)}}{d\tau}\right)  ^{2}\right]
\label{IX-20-A}\\
& -\frac{1}{4}\left[  \frac{d\ln\Gamma_{(1)}}{d\tau}+\frac{d\ln\Gamma_{(2)}%
}{d\tau}+\frac{d\ln\Gamma_{(3)}}{d\tau}\right]  ^{2}\nonumber\\
& +\frac{1}{2}I_{\left(  1\right)  }\left(  \Omega_{\left(  1\right)
}\right)  ^{2}+\frac{1}{2}I_{\left(  2\right)  }\left(  \Omega_{\left(
2\right)  }\right)  ^{2}+\frac{1}{2}I_{\left(  3\right)  }\left(
\Omega_{\left(  3\right)  }\right)  ^{2}\nonumber\\
& -\frac{1}{2}\left[  \left(  \Gamma_{(1)}\right)  ^{2}+\left(  \Gamma
_{(2)}\right)  ^{2}+\left(  \Gamma_{(3)}\right)  ^{2}-2\Gamma_{(1)}%
\Gamma_{(2)}-2\Gamma_{(1)}\Gamma_{(3)}-2\Gamma_{(2)}\Gamma_{(3)}\right]
.\nonumber
\end{align}
Direct calculation shows that equations of motion (\ref{IX-15})--(\ref{IX-19})
indeed follow from this $\mathcal{L}$ as the standard Lagrange equations if we
identify $\ln\Gamma_{(1)},$ $\ln\Gamma_{(2)},$ $\ln\Gamma_{(3)},$ $\psi,$
$\theta,$ $\varphi$ with generalized coordinates and their time derivatives
with respect to $\tau$ with generalized velocities. It is also easy to show
that this Lagrangian coincides with the one which follows in the standard way
from the Einstein-Hilbert action written in the synchronous system and with
respect to the time $\tau$. The integrand in such an action, after discarding
a total time derivative, gives (up to a constant multiplier) the right hand
side of expression (\ref{IX-20-A}).

Let's look again at Eqs.~(\ref{IX-15})--(\ref{IX-20}) for the general Bianchi
type IX model. These can be simplified due to the fact that the Euler
equations (\ref{IX-15}) have three exact integrals of the motion which without
loss of generality can be reduced to the following relations:
\begin{equation}
I_{\left(  1\right)  }\Omega_{\left(  1\right)  }=J\sin\theta\sin\psi\text{
},\text{ }I_{\left(  2\right)  }\Omega_{\left(  2\right)  }=J\sin\theta
\cos\psi\text{ },\text{ }I_{\left(  3\right)  }\Omega_{\left(  3\right)
}=J\cos\theta,\label{IX-22}%
\end{equation}
where $J$ is an arbitrary constant (two other arbitrary constants of
integration can be eliminated due to the gauge freedom of rotations of the
vectors $l_{\alpha}^{\left(  a\right)  }$, which rotations do not change the
already chosen canonical set of the structure constants $l_{(a)}^{\mu}%
l_{(b)}^{\nu}[l_{\mu,\nu}^{(c)}-l_{\nu,\mu}^{(c)}]$).

Eqs.~(\ref{IX-22}), (\ref{IX-17})--(\ref{IX-20}) together with the definitions
(\ref{IX-12}) and (\ref{IX-11}) of the \textquotedblleft moments of inertia"
and ``angular velocities" constitute a self-consistent and closed system of
equations for the six dynamic variables $\theta,\psi,\varphi,\Gamma
_{(1)},\Gamma_{(2)},\Gamma_{(3)}$ governing the evolution of the general
homogeneous models of type IX. This system, in spite of its homogeneity, is
still too complicated to be resolved analytically but it turns out that in the
asymptotic vicinity of the cosmological singularity it simplifies drastically.
The point is that near the singularity the time dependence of the matrix
$O_{\left(  b\right)  }^{\left(  a\right)  }$ terminates. The scale factors
$\Gamma_{\left(  a\right)  }$ continue to oscillate but they never cross each
other which can be seen from their dynamical equations (\ref{IX-17}%
)--(\ref{IX-19}). These equations contain terms which become singular when any
two of the scale factors coincide. These singularities reflect the unlimited
growth of centrifugal forces which do not permit any two of the scale factors
to approach each other too closely. Then what happens is the following: if at
some instant they are arranged,\ for example, in the order $\Gamma
_{(1)}>\Gamma_{(2)}>\Gamma_{(3)}$ then this order is maintained from then on
and near the singularity the inequalities grow until
\begin{equation}
\Gamma_{(1)}\gg\Gamma_{(2)}\gg\Gamma_{(3)},\label{Hom48}%
\end{equation}
that is the anisotropy of space grows without bound. This inequality means
that the ratios $\Gamma_{(2)}/\Gamma_{(1)},$ $\Gamma_{(3)}/\Gamma_{(1)}$ and
$\Gamma_{(3)}/\Gamma_{(2)}$ go to zero:%
\begin{equation}
\Gamma_{(2)}/\Gamma_{(1)}\rightarrow0,\text{ }\Gamma_{(3)}/\Gamma
_{(1)}\rightarrow0,\text{\ }\Gamma_{(3)}/\Gamma_{(2)}\rightarrow
0.\label{Hom49}%
\end{equation}
From this it follows that all three \textquotedblleft moments of inertia"
(\ref{IX-12}) tend to infinity and equations (\ref{IX-22}) show that all three
angular velocities $\Omega_{\bar{\alpha}}$ go to zero near the singularity.
This leads to the result that the Euler angles approach three arbitrary
constants (in the further inhomogeneous generalization these will be three
arbitrary 3-dimensional functions):
\begin{equation}
\left(  \theta,\varphi,\psi\right)  \rightarrow\left(  \theta_{0},\varphi
_{0},\psi_{0}\right)  \,.\label{Hom50}%
\end{equation}

Indeed, suppose that near the singularity $\tau\rightarrow\infty$ the Euler
angles tend to the limits (\ref{Hom50}). Let's introduce the notation:%
\begin{equation}
\Gamma_{(1)}\equiv A^{2},\text{ }\Gamma_{(2)}J^{2}\cos^{2}\theta_{0}\equiv
B^{2},\text{ }\Gamma_{(3)}J^{4}\sin^{2}\theta_{0}\cos^{2}\theta_{0}\sin
^{2}\psi_{0}\equiv C^{2}.\label{Hom51}%
\end{equation}
From Eqs. (\ref{IX-17})--(\ref{IX-20}) under the asymptotic conditions
(\ref{Hom49}) we obtain the following equations, where we have kept only the
leading terms:
\begin{align}
& 2\frac{\partial^{2}\ln A}{\partial\tau^{2}}=\frac{B^{2}}{A^{2}}%
-A^{4},\label{Hom52}\\
&  2\frac{\partial^{2}\ln B}{\partial\tau^{2}}=A^{4}-\frac{B^{2}}{A^{2}}%
+\frac{C^{2}}{B^{2}},\label{Hom53}\\
&  2\frac{\partial^{2}\ln C}{\partial\tau^{2}}=A^{4}-\frac{C^{2}}{B^{2}%
},\label{Hom54}%
\end{align}
\begin{gather}
\left(  \frac{\partial\ln A}{\partial\tau}\right)  ^{2}+\left(  \frac
{\partial\ln B}{\partial\tau}\right)  ^{2}+\left(  \frac{\partial\ln
C}{\partial\tau}\right)  ^{2}-\left(  \frac{\partial\ln A}{\partial\tau}%
+\frac{\partial\ln B}{\partial\tau}+\frac{\partial\ln C}{\partial\tau}\right)
^{2}\label{Hom55}\\
=-\frac{1}{2}\left(  A^{4}+\frac{B^{2}}{A^{2}}+\frac{C^{2}}{B^{2}}\right)
.\nonumber
\end{gather}
Note that also in this approximation Eq.~(\ref{Hom55}) can be shown to be an
exact integral of Eqs.~(\ref{Hom52})--(\ref{Hom54}).

Now it is easy to verify (\textit{post factum}) the asymptotics (\ref{Hom48}%
)--(\ref{Hom49}) by an analysis of Eqs.~(\ref{Hom52})--(\ref{Hom55}) of the
same kind that we used above for an analysis of the process of changing Kasner
epochs. The solution of this system can be studied by dividing the evolution
again into epochs during each of which the right-hand sides of
Eqs.~(\ref{Hom52})--(\ref{Hom55}) can be neglected. For each epoch we obtain
again the Kasner solution for the functions $A^{2},B^{2},C^{2}$ and in the
right hand side of Eqs.~(\ref{Hom52})--(\ref{Hom55}) again will appear
``dangerous" terms which will grow in the course of approaching the
singularity and which sooner or later will destroy this Kasner regime. Each
time we will have only one dominant growing term among $A^{4},$ $B^{2}/A^{2},
$ $C^{2}/B^{2}$ and each time transition to the new regime can be studied
separately under the influence of this term. The solution of Eqs.~(\ref{Hom52}%
)--(\ref{Hom55}) on the right hand side of which we keep only one of these
``dangerous" terms shows that each of them begins to grow, reach a maximum,
and after a relatively short time dies away and their influence once again
becomes small. If at the initial epoch the growing function $A^{4} $ dominates
then the corresponding transition is exactly the same we already described in
the first section. We call this already known transition a
\textit{gravitational bounce}. If the growing dominant term is $B^{2}/A^{2}$
(or $C^{2}/B^{2}$) we will have a new type of transition we will call a
\textit{centrifugal bounce.} Again, the term $B^{2}/A^{2}$ (or $C^{2}/B^{2}$)
begins to grow, reach a maximum, and after some time dies away.

Indeed consider Eqs.~(\ref{Hom52})--(\ref{Hom55}) with only $B^{2}/A^{2}$ on
the right-hand side (the action of $C^{2}/B^{2}$ has exactly the same
character). The exact general solution of the system in this case shows that
before the transition ($\tau\rightarrow-\infty,$ $t\rightarrow\infty)$ the
asymptotic behavior of the solution, transformed to the synchronous time $t,$
is $\left(  A^{2},B^{2},C^{2}\right)  \sim\left(  t^{2p_{A}},t^{2p_{B}%
},t^{2p_{C}}\right)  ,$ and after the transition ($\tau\rightarrow\infty,$
$t\rightarrow0$) we have $\left(  A^{2},B^{2},C^{2}\right)  =\left(
t^{2p_{B}},t^{2p_{A}},t^{2p_{C}}\right)  .$ We see that in the course of this
centrifugal transition the numerical values of the exponents do not change.
Two of the three Kasner exponents simply change places. A description of the
transformation under the influence of $C^{2}/B^{2}$ is obtained from the above
formulas by the following permutation of the components $\left(  A^{2}%
,B^{2},C^{2}\right)  \rightarrow\left(  B^{2},C^{2},A^{2}\right)  .$

We find an endless succession of the Kasner regimes between each of which
there is one of two types of transformations, one due to the growth of the
perturbation $A^{4}$ and another due to the perturbation $B^{2}/A^{2}$ or
$C^{2}/B^{2}$. Also it is easy to show that as we approach the singularity
($\tau\rightarrow\infty,$ $t\rightarrow0$) all three quantities $A^{4}%
,B^{2}/A^{2},C^{2}/B^{2}$ go to zero because their maxima go to zero. The
foregoing arguments prove the asymptotic inequalities (\ref{Hom48}%
)--(\ref{Hom49}).

As for the limit (\ref{Hom50}) for the Euler angles we remark that
Eqs.~(\ref{IX-22}) can be written as%
\begin{equation}
\frac{\partial\varphi}{\partial\tau}\sin\theta\sin\psi+\frac{\partial\theta
}{\partial\tau}\cos\psi=\frac{J\Gamma_{(2)}\Gamma_{(3)}\sin\theta\sin\psi
}{\left(  \Gamma_{(2)}-\Gamma_{(3)}\right)  ^{2}},\label{Hom41}%
\end{equation}%
\begin{equation}
\frac{\partial\varphi}{\partial\tau}\sin\theta\cos\psi-\frac{\partial\theta
}{\partial\tau}\sin\psi=\frac{J\Gamma_{(1)}\Gamma_{(3)}\sin\theta\cos\psi
}{\left(  \Gamma_{(1)}-\Gamma_{(3)}\right)  ^{2}},\label{Hom42}%
\end{equation}%
\begin{equation}
\frac{\partial\varphi}{\partial\tau}\cos\theta+\frac{\partial\psi}%
{\partial\tau}=\frac{J\Gamma_{(1)}\Gamma_{(2)}\cos\theta}{\left(  \Gamma
_{(1)}-\Gamma_{(2)}\right)  ^{2}},\label{Hom43}%
\end{equation}
and can be used to find the corrections $\delta\theta,\delta\varphi,\delta
\psi$ to the first approximation $\left(  \theta,\varphi,\psi\right)  =\left(
\theta_{0},\varphi_{0},\psi_{0}\right)  .$ Writing%
\begin{equation}
\theta=\theta_{0}+\delta\theta,\text{ \ }\varphi=\varphi_{0}+\delta
\varphi,\text{ \ }\psi=\psi_{0}+\delta\psi,\label{Hom56}%
\end{equation}
we find that in the first approximation from (\ref{Hom41})--(\ref{Hom43})
follows:%
\begin{equation}
\frac{\partial}{\partial\tau}\delta\theta=\frac{C^{2}}{B^{2}}\frac{\cos
\psi_{0}}{J\sin\theta_{0}\sin\psi_{0}}\label{Hom57}%
\end{equation}%
\begin{equation}
\frac{\partial}{\partial\tau}\delta\varphi=\frac{C^{2}}{B^{2}}\frac{1}%
{J\sin^{2}\theta_{0}}\label{Ho58}%
\end{equation}%
\begin{equation}
\frac{\partial}{\partial\tau}\delta\psi=\left(  \frac{B^{2}}{A^{2}}\frac
{1}{\cos^{2}\theta_{0}}-\frac{C^{2}}{B^{2}}\frac{1}{\sin^{2}\theta_{0}%
}\right)  \frac{\cos\theta_{0}}{J}\label{Hom59}%
\end{equation}

Now it can be confirmed that the fact that all three derivatives
$\frac{\partial\psi}{\partial\tau},\frac{\partial\theta}{\partial\tau}%
,\frac{\partial\varphi}{\partial\tau}$ in the limit $\tau\rightarrow\infty$ go
to zero (due to the inequalities (\ref{Hom48})--(\ref{Hom49})) actually leads
to the existence of the limits given in (\ref{Hom50}) for the angles
$\psi,\theta,\varphi.$ The corresponding analysis can be found in
Ref.~\refcite{BKR}. The phenomenon of cessation of rotation in the aymptotic
vicinity of the singularity we call \textit{freezing} (terminology proposed by
DHN\cite{DHN}).

\section{Cosmological billiard for the general Bianchi type IX model}

The oscillatory asymptotics following from Eqs.~(\ref{Hom52})--(\ref{Hom55})
again can be described as endless oscillations of a particle against the walls
of some potential in the 3-dimensional $\alpha$-space analogous to the $\beta
$-space we used in the section \textquotedblleft Cosmological billiard" for
the diagonal Bianchi type IX model. Eqs.~(\ref{Hom52})--(\ref{Hom54}) after
introducing the variables analogous to (\ref{OR3-5})%
\begin{equation}
A^{2}=e^{-2\alpha^{\left(  1\right)  }},\text{ }B^{2}=e^{-2\alpha^{\left(
2\right)  }},\text{ }C^{2}=e^{-2\alpha^{\left(  3\right)  }},\label{Hom112}%
\end{equation}
follow from the Lagrangian:%

\begin{equation}
L=G_{\left(  a\right)  \left(  b\right)  }\frac{\partial\alpha^{\left(
a\right)  }}{\partial\tau}\frac{\partial\alpha^{\left(  b\right)  }}%
{\partial\tau}-V(\alpha),\label{Hom115}%
\end{equation}

\begin{equation}
V(\alpha)=\frac{1}{2}\left[  e^{-4\alpha^{\left(  1\right)  }}+e^{-2\left(
\alpha^{\left(  2\right)  }-\alpha^{\left(  1\right)  }\right)  }+e^{-2\left(
\alpha^{\left(  3\right)  }-\alpha^{\left(  2\right)  }\right)  }\right]
,\label{Hom116}%
\end{equation}
where $G_{\left(  a\right)  \left(  b\right)  }$ is defined by the same
formula (\ref{OR3-11}) in which the letter $\beta$ should be replaced by
$\alpha$. The additional \textquotedblleft energy" constraint (\ref{Hom55})
has the same form:%
\begin{equation}
G_{\left(  a\right)  \left(  b\right)  }\frac{\partial\alpha^{\left(
a\right)  }}{\partial\tau}\frac{\partial\alpha^{\left(  b\right)  }}%
{\partial\tau}+V(\alpha)=0.\label{Hom117}%
\end{equation}
In the preceding section we showed that near the singularity ($\tau
\rightarrow\infty,$ $t\rightarrow0$) all three quantities $A^{2},B^{2}%
/A^{2},C^{2}/B^{2}$ go to zero which means that all three exponents
$\alpha^{\left(  1\right)  },$ $\alpha^{\left(  2\right)  }-\alpha^{\left(
1\right)  },$ $\alpha^{\left(  3\right)  }-\alpha^{\left(  2\right)  }$ tend
to plus infinity. Consequently, in this limit the potential vanishes
everywhere except in the narrow regions near the reflecting walls
$\alpha^{\left(  1\right)  }=0,$ $\alpha^{\left(  2\right)  }-\alpha^{\left(
1\right)  }=0,$ $\alpha^{\left(  3\right)  }-\alpha^{\left(  2\right)  }=0.$
The first of these is the same \textit{gravitational wall} as in the case of
non-rotating axes but the other two represent the new \textit{centrifugal
walls. }If we look to the exact potential written out in the Lagrangian
(\ref{IX-20-A}) we see that the old walls $\alpha^{\left(  2\right)  }=0$ and
$\alpha^{\left(  3\right)  }=0$, of course, still remain in the system but
near to the singularity they ceased to be dominant, that is they have no more
influence on the dynamics in the first approximation. Instead the new
centrifugal walls $\alpha^{\left(  2\right)  }-\alpha^{\left(  1\right)  }=0,$
$\alpha^{\left(  3\right)  }-\alpha^{\left(  2\right)  }=0$ became dominant,
pressing back the old two behind them. Then the new billiard table occupies
now only 1/6 part of the volume of the old triangle (the old triangle we
should cut by the three bisectrices which are the new centrifugal walls and
take only one region bounded by the one old gravitational wall and by two new
centrifugal walls).

It turns out that also this new shortened part (in spite of the fact that its
two new reflecting borders are different) of the old billiard table possesses
the same four basic properties we enumerated at the end of the section
\textquotedblleft Cosmological billiard". The billiard for the general Bianchi
type IX model with rotating axes described in the present section also has the
Kac-Moody symmetry but of a different type with respect to the one we
discussed in section \textquotedblleft First manifestation of the hidden
Kac-Moody algebra". Now the fundamental Weyl chamber in $\alpha$- space has
different boundary walls: $\alpha^{\left(  1\right)  }=0,$ $\alpha^{\left(
2\right)  }-\alpha^{\left(  1\right)  }=0,$ $\alpha^{\left(  3\right)
}-\alpha^{\left(  2\right)  }=0$ and the vectors orthogonal to these walls are
also different. These vectors we designate again by $w_{A}$ and in components
they are: $w_{1}=(1,0,0),$ $w_{2}=(-1/2,1/2,0),$ $w_{3}=(0,-1/2,1/2)$. The
scalar products $(w_{A}\bullet w_{B})=G^{\left(  a\right)  \left(  b\right)
}w_{A\left(  a\right)  }w_{B\left(  b\right)  }$ we again can arrange in the
form of the matrix (\ref{Kac1}) and calculation gives:%
\begin{equation}
A_{AB}=\left(
\begin{array}
[c]{ccc}%
2 & -2 & 0\\
-2 & 2 & -1\\
0 & -1 & 2
\end{array}
\right)  .\label{Kac4}%
\end{equation}
This is again a Cartan matrix of indefinite type which corresponds to one of
the infinite-dimensional Lorenzian hyperbolic Kac-Moody algebras of rank 3.
This algebra has a variety of names depending on the authors who introduced
it. We call it $AE_{3}$ as in Ref.~\refcite{DHN} (it also called $H_{3}$ in
Ref.~\refcite{Kac}, $A_{1}^{\wedge\wedge}$ in Ref.~\refcite{DHJN}) and
$A_{1}^{++}$ in Ref.~\refcite{HPS}).

For the homogeneous model of Bianchi type IX the approach which uses the
orthogonal matrix $O_{\left(  b\right)  }^{\left(  a\right)  }$ has been
developed in Ref.~\refcite{BKR}. Although the corresponding generalization to
the inhomogeneous case is straightforward, we never did it. However, the
recent development of the theory has shown that even in most general
inhomogeneous cases (including multidimensional space-time filled by different
kinds of matter) there is another representation of the metric tensor leading
to the same asymptotic freezing phenomenon of \textquotedblleft non-diagonal"
degrees of freedom and reducing the full dynamics to the few \textquotedblleft
diagonal" oscillating scale factors which can be described in the same way as
the cosmological billiard. This is the so-called Iwasawa (triangular)
decomposition of the metric tensor first used by DHN \cite{DHN} and thoroughly
investigated in Ref.~\refcite{DB}. The point is that in the general
inhomogeneous case it is convenient to use an upper triangular matrix $N$
\ (with components $N_{\alpha}^{\left(  a\right)  }$ where again the upper
index numerates the rows and lower index corresponds to columns)%

\begin{equation}
N_{\alpha}^{\left(  a\right)  }=\left(
\begin{array}
[c]{ccc}%
1 & n_{1} & n_{2}\\
0 & 1 & n_{3}\\
0 & 0 & 1
\end{array}
\right)  \text{\ \ }\label{3}%
\end{equation}
and to write the line element in the form $g_{\alpha\beta}dx^{\alpha}%
dx^{\beta}=D_{\left(  a)(b\right)  }N_{\alpha}^{\left(  a\right)  }N_{\beta
}^{\left(  b\right)  }dx^{\alpha}dx^{\beta}).$ The diagonal matrix $D$ as well
as the matrix $N$ are functions of all four coordinates but near the
singularity the matrix $N$ tends to some time-independent limit and the
components of $D$ oscillate between the walls of a potential of the same
structure as (\ref{Hom116}).

\section{The influence of matter}

In Refs.~\refcite{BK3,BK4,BK5} we studied the problem of the influence of
various kinds of matter upon the behavior of the general inhomogeneous
solution of the gravitational field equations in the neighborhood of the
singular point. It is clear that, depending on the form of the energy-momentum
tensor, we may meet three different possibilities: (i) the oscillatory regime
remains as it is in vacuum, i.e., the influence of matter may be ignored in
the first approximation; (ii) the presence of matter makes the existence of
Kasner epochs near a singular point impossible; (iii) Kasner epochs exist as
before, but matter affects the process of their formation and alternation.
Actually, all these possibilities may be realized.

There is a case in which the oscillatory regime observed as a singular point
is approached remains the same, in the first approximation, as in vacuum. This
case is realized in a space filled with a perfect fluid with the equation of
state $p=k\varepsilon$ for $0\leqslant k<1$. No additional reflecting walls
arise from the energy-momentum tensor in this case.

If $k=1$ we have the ``stiff matter" equation of state $p=\varepsilon$. This
is an example of the second of the above-mentioned possibilities when neither
Kasner epoch nor oscillatory regime can exist in the vicinity of a singular
point. This case has been investigated in Refs.~\refcite{BK3, BK4} where it
has been shown that the influence of the ``stiff matter" (equivalent to a
massless scalar field) results in the violation of the Kasner relations
(\ref{OR2}) for the asymptotic exponents. Instead we have%
\begin{equation}
p_{\left(  1\right)  }+p_{\left(  2\right)  }+p_{\left(  3\right)  }=1,\text{
\ \ \ \ }p_{\left(  1\right)  }^{2}+p_{\left(  2\right)  }^{2}+p_{\left(
3\right)  }^{2}=1-p_{\varphi}^{2}\,,\label{4}%
\end{equation}
where $p_{\varphi}^{2}$ is an arbitrary three-dimensional function (with the
restriction $p_{\varphi}^{2}<1$) to which the energy density $\varepsilon$ of
the matter is proportional (in that particular case when the stiff matter
source is realized as a massless scalar field $\varphi$ its asymptotic is
$\varphi=p_{\varphi}\ln t$ and this is the formal reason why we use the index
$\varphi$ for the additional exponent $p_{\varphi}$).

Thanks to (\ref{4}), in contrast to the Kasner relations (\ref{OR2}), it is
possible for all three exponents $p_{\left(  a\right)  }$ to be positive. In
Ref.~\refcite{BK4} it has been shown that, even if the contraction of space
starts with the quasi-Kasner epoch (\ref{4}) during which one of the exponents
$p_{\left(  a\right)  }$ is negative, the power law asymptotic behavior with
all positive exponents is inevitably established after a finite number of
oscillations and remains unchanged up to the singular point. Thus, for the
equation of state $p=\varepsilon$ the collapse in the general solution is
described by monotonic (but anisotropic) contraction of space along all
directions. We constructed the asymptotics of the general solution near the
cosmological singularity for this case explicitly in Ref.~\refcite{BK3}, see
also Refs.~\refcite{BK4, AR, DHRW}. The disappearance of oscillations for the
case of a massless scalar field should be considered as an isolated phenomenon
which is unstable with respect to inclusion into the right hand side of the
Einstein equations with another kinds of fields. For instance, in the same
paper \cite{BK4} we showed that if we add a vector field to the scalar field
then the endless oscillations reappear.

Another example of the generic solution with smooth power law asymptotics can
be constructed for the viscoelastic material with shear viscosity coefficient
$\eta$ which near the singularity behaves as $\eta\sim\sqrt{\varepsilon}$.
This kind of matter can stabilize the Friedmann cosmological singularity, and
the corresponding generic solution of the Einstein equations possessing an
isotropic big bang (or big crunch) exists \cite{B}.

The cosmological evolution in the presence of an electromagnetic field may
serve as an example of the third possibility. In this case the oscillatory
regime in the presence of matter is, as usual, described by the alternation of
Kasner epochs, but in this process the energy-momentum tensor plays a role as
important as the three-dimensional curvature tensor. This problem has been
treated by us in Ref.~\refcite{BK5}, where it has been shown that in addition
to the vacuum reflecting walls, there are also new walls which arise from the
energy-momentum tensor of the electromagnetic field.

In Ref.~\refcite{BK3} we have also studied the problem of the influence of
Yang-Mills fields on the character of the cosmological singularity. For
definiteness, we have restricted ourselves to fields corresponding to the
gauge group SU(2). The study was performed in the synchronous reference system
in the gauge in which the time components of all three vector fields are equal
to zero. It was shown that, in the neighborhood of a cosmological singularity,
the behavior of the Yang-Mills fields is largely similar to the behavior of
the electromagnetic field: as before, there appears an oscillatory regime
described by the alternation of Kasner epochs, which is caused either by the
three-dimensional curvature or by the energy-momentum tensor. If, in the
process of the alternation of epochs, the energy-momentum tensor of the gauge
fields is dominant, the qualitative behavior of the solution during the epochs
and in the transition regions between them is like the behavior in the case of
a free Yang-Mills field (with an Abelian symmetry group). This does not mean
that non-linear terms of the interaction may be neglected completely, but the
latter introduce only minor quantitative changes into the picture we would
observe in the case of non-interacting fields.

\section{On supergravity}

An interesting fact is that the bosonic sectors of supergravities emerging in
the low energy limit from all types of superstring models have an oscillatory
cosmological singularity of the BKL character \cite{DH1, DH2}. To clarify the
main points consider an action of the following general form:%
\begin{equation}
S=\int d^{D}x\sqrt{g}\text{ }\left[  R-\partial^{i}\varphi\partial_{i}%
\varphi-\frac{1}{2}\sum_{p}\frac{1}{(p+1)!}e^{\lambda_{p}\varphi}%
F_{i_{1}...i_{p+1}}^{(p+1)}F^{(p+1)i_{1}...i_{p+1}}\right]  ,\label{5}%
\end{equation}
where $F^{(p+1)}$ designates the field strengths generated by the $p$-forms
$A_{p}$, i.e., $F_{i_{1}...i_{p+1}}^{(p+1)}=antisym(\partial_{i_{1}}%
A_{i_{2}...i_{p+1}})$. The real parameters $\lambda_{p}$ are coupling
constants corresponding to the interaction between the dilaton and $p$-forms.
The tensorial operations in (\ref{5}) are carried out with respect to the
$D$-dimensional metric $g_{ik}$ and $g=\left\vert \det g_{ik}\right\vert $.
Now the lowercase Latin letters refer to the tensorial indices of
$D$-dimensional space-time and Greek indices correspond to the $d$-dimensional
tensorial indices, where $d=D-1.$ We will use the two lowercase Latin indices
in brackets $\left(  a\right)  $ and $\left(  b\right)  $ as before to
designate the frame indices of the $d$-dimensional space. Also in this theory
generalized Kasner-like epochs exist which are of the form:
\begin{equation}
g_{ik}dx^{i}dx^{k}=-dt^{2}+\eta_{\left(  a\right)  \left(  b\right)
}l_{\alpha}^{\left(  a\right)  }l_{\beta}^{\left(  b\right)  }dx^{\alpha
}dx^{\beta},\label{6}%
\end{equation}%
\begin{equation}
\eta_{\left(  a\right)  \left(  b\right)  }=diag[t^{2p_{\left(  1\right)  }%
},t^{2p_{\left(  2\right)  }},...,t^{2p_{\left(  d\right)  }}],\label{7}%
\end{equation}%
\begin{equation}
\varphi=p_{\varphi}\ln t+\varphi_{0},\label{8}%
\end{equation}
where $\eta_{\left(  a\right)  \left(  b\right)  },l_{\alpha}^{\left(
a\right)  },p_{\left(  a\right)  },p_{\varphi},\varphi_{0}$ are functions on
space coordinates $x^{\alpha}.$ In the presence of the dilaton the exponents
$p_{\left(  a\right)  \text{ }}$and $p_{\varphi}$ instead of the Kasner law
satisfy the relations analogous to (\ref{4}):
\begin{equation}
\sum_{a=1}^{d}p_{\left(  a\right)  }=1,\text{ \ \ }\sum_{a=1}^{d}p_{\left(
a\right)  }^{2}{}=1-p_{\varphi}^{2}\label{19}%
\end{equation}

The approximate solution (\ref{6})--(\ref{8}) follows from the $D$-dimensional
Einstein equations by neglecting the energy-momentum tensor of $p$-forms,
$d$-dimensional curvature tensor $P_{\alpha\beta}$ and spatial derivatives of
$\varphi$. Now one has to do the work analogous to that one in 4-dimensional
gravity: it is necessary to identify in all neglected parts of the equations
those \textquotedblleft dangerous" terms which destroy the solution
(\ref{6})--(\ref{8}) in the limit $t\rightarrow0$. Then one should construct
the new first approximation to the equations taking into account also these
\textquotedblleft dangerous\textquotedblright\ terms and try to find the
asymptotic solution for this new system. This is the same method which was
used in the case of 4-dimensional gravity and it also works well here. Using
the Iwasawa decomposition for the $d$-dimensional metric $g_{\alpha\beta
}=D_{\left(  a)(b\right)  }N_{\alpha}^{\left(  a\right)  }N_{\beta}^{\left(
b\right)  }$ where $D_{\left(  a)(b\right)  }=\mathrm{diag}(e^{-2\alpha
^{\left(  1\right)  }},e^{-2\alpha^{\left(  2\right)  }},...,e^{-2\alpha
^{\left(  d\right)  }})$ it can be shown \cite{DHN} that near the singularity
again the phenomenon of freezing of the \textquotedblleft non-diagonal"
degrees of freedom of the metric tensor arises (that is the matrix $N_{\alpha
}^{\left(  a\right)  }$ becomes time-independent) and the system reduces to
ordinary differential equations (for each spatial point) in time for the
variables $\alpha^{\left(  1\right)  },...,\alpha^{\left(  d\right)  }$ and
$\varphi$. It is convenient to use the $(d+1)$-dimensional flat superspace
with coordinates $\alpha^{\left(  1\right)  },...,\alpha^{(d)},\varphi$. The
asymptotic dynamics for these variables follows from the Lagrangian of the
form similar to (\ref{Hom115}):%
\begin{equation}
L=G_{\left(  a\right)  \left(  b\right)  }\frac{\partial\alpha^{\left(
a\right)  }}{\partial\tau}\frac{\partial\alpha^{\left(  b\right)  }}%
{\partial\tau}+\left(  \frac{d\varphi}{d\tau}\right)  ^{2}-\sum\limits_{q}%
C_{q}e^{-2W_{q}\left(  \alpha\right)  },\text{ \ \ }\label{20}%
\end{equation}
where $G_{\left(  a\right)  \left(  b\right)  }$ is defined again by the
formula (\ref{OR3-11}) in which the letter $\beta$ should be replaced by
$\alpha$ and the time variable $\tau$ relates to the synchronous time $t$ by
the formula $dt=-\sqrt{\det g_{\alpha\beta}}\,d\tau$. The $00-$component of
the Einstein equations gives the additional condition to the solutions of the
equations of motion:
\begin{equation}
G_{\left(  a\right)  \left(  b\right)  }\frac{\partial\alpha^{\left(
a\right)  }}{\partial\tau}\frac{\partial\alpha^{\left(  b\right)  }}%
{\partial\tau}+\left(  \frac{d\varphi}{d\tau}\right)  ^{2}+\sum\limits_{q}%
C_{q}e^{-2W_{q}\left(  \alpha\right)  }=0.\label{21}%
\end{equation}
The sum in the potential means summation over all relevant (dominating)
impenetrable walls (numerated by the index $q$) located at hypersurfaces where
$W_{q}=0$. All functional parameters $C_{q}$ (depending on the coordinates
$x^{\alpha}$) in general are positive.

The free motion of a particle between the walls in the original $(d+1)$%
-dimensional superspace with coordinates $\alpha^{\left(  1\right)
},...,\alpha^{(d)},\varphi$ is projected onto geodesic motion on the
hyperbolic $d$-dimensional $\gamma$-space by the same type of the projection
used in 4-dimensional gravity:%
\begin{equation}
\alpha_{\text{ }}^{\left(  a\right)  }=\rho\gamma^{\left(  a\right)  },\text{
}\varphi=-\rho\gamma^{\left(  d+1\right)  },\ G_{\left(  a)\left(  b\right)
\right)  }\gamma_{\text{ }}^{\left(  a\right)  }\gamma_{\text{ }}^{\left(
b\right)  }+\left[  \gamma^{\left(  d+1\right)  }\right]  ^{2}=-1,\text{ }%
\rho=const.,\label{22}%
\end{equation}
that is to motion between the corresponding projections of the original walls
$W_{q}=0$ onto $\gamma$-space. These projections bound a region in $\gamma
$-space inside of which a symbolic particle oscillates and the volume of this
region, in spite of its non-compactness, is finite. The last property is of
principle significance since it leads, as in the 4-dimensional case, to the
chaotic character of the oscillatory regime.

Of course, the central point here is to find all the aforementioned dominant
walls and corresponding \textquotedblleft wall forms" $W_{q}(\alpha).$ This
depends on the space-time dimension and menu of $p$-forms. In
Refs.~\refcite{DH3, DHN} the detailed description of all the possibilities for
all types of supergravities (i.e., eleven-dimensional supergravity and those
following from the known five types of superstring models in ten-dimensional
space-time) can be found. It was shown that in all cases there are only 10
relevant walls governing the oscillatory dynamics. The huge number of other
walls need not be considered because they are located behind these principal
ten and have no influence on the dynamics in the first approximation. All
dihedral angles between the dominant walls are equal to the numbers $\pi/n$
where $n$ belongs to some distinguished set of natural numbers (or equal to
infinity). This is exactly the geometrical structure which generates the set
of 10 simple roots (vectors normal to walls) of an infinite-dimensional
hyperbolic Kac-Moody algebra of rank 10.

The above-mentioned region in $\gamma$-space represents the multidimensional
``billiard table" and the collection of its bounding cushions forms the
generalization to the covering space of the so-called Coxeter crystallographic
simplex, that is, in the cases under consideration, a polyhedron with 10 faces
in 9-dimensional $\gamma$-space with special angles between the cushions.

The fermionic sectors of supergravities have been thoroughly investigated by a
number of authors which led to the construction of a \textquotedblleft
super-billiard" having both bosonic and fermionic degrees of freedom. An
interested reader can look at the recent paper Ref. \refcite{DS} dedicated to
this problem where one can also find the basic references in relation to the
fermionic aspect of the theory of the oscillatory cosmological singularity.
\bigskip

\section{Acknowledgment}

It is the pleasure to thank Robert Jantzen for the editing work, useful
comments and improvement of English.

\bigskip\ \qquad

\bigskip

\end{document}